\documentclass[journal,transmag]{IEEEtran}
\usepackage{amsfonts}
\usepackage{bbm}
\usepackage{subfigure}
\usepackage{amsmath}
\usepackage{epsfig}
\usepackage{amssymb}
\usepackage{mathrsfs}
\usepackage{algorithmic}
\usepackage[ruled]{algorithm}
\usepackage{footmisc}
\usepackage{xspace}
\usepackage{array}
\usepackage{multirow}
\usepackage{graphicx}

\newtheorem{theorem}{Theorem}

\newtheorem{lemma}[theorem]{Lemma}

\def\ie{\textit{i.e.}\xspace}

\def\alg{AVA\xspace}
\def\arm{a}
\def\armnum{K}                                   
\def\charge{\lambda}                         
\def\cost{c}
\def\Energy{E}                                      
\def\energy{e}                                      
\def\energyTh{\energy_t}
\def\expect{\mathbb{E}}
\def\node{v}
\def\paraa{\alpha}
\def\parab{\beta}
\def\parac{\gamma}
\def\parad{\delta}
\def\parae{\varepsilon}
\def\period{T}                                        
\def\prob{p}                                           
\def\protocol{ODC\xspace}
\def\pullarmnumber{\varphi\xspace}
\def\pullarmnumtotal{\Psi\xspace}
\def\pullvariable{\chi\xspace}
\def\remaindata{M}
\def\reward{I}
\def\rewardth{\reward_d}                     
\def\rewarde{\bar{\reward}}                
\def\rewardd{\Delta}                             
\def\regret{R}
\def\route{{\L}}
\def\scheme{X}
\def\schemeO{COA\xspace}
\def\schemeS{SDC\xspace}
\def\slot{\tau}
\def\set{s}
\def\solarstate{\theta}
\def\VoIbound{\hat{\reward}}                   
 \def\vectorpara{A}
\def\vectorfeature{B}

\include{figures}
 \allowdisplaybreaks 

\begin{document}
%
\title{Value of Information Aware Opportunistic Duty Cycling in Solar Harvesting Sensor Networks }
\author{\IEEEauthorblockN{Jianhui Zhang}
\IEEEauthorblockA{College of Computer Science and Technology, Hangzhou Dianzi University, 310018 China.\\
Email: jhzhang@ieee.org}
}

\maketitle
\begin{abstract}
The energy-harvested Wireless Sensor Networks (WSNs) may operate perpetually with the extra energy supply from ambient natural energy, such as solar energy.
Nevertheless, the harvested energy is still limited so it's not able to support the perpetual network operation with full duty cycle.
To achieve the perpetual network operation and process the data with high importance, measured by Value of Information (VoI), the network has to operate under partial duty cycle and to improve the  efficiency to consume the harvested energy.
The challenging problem is how to deal with the stochastic feature of the natural energy and the variable data VoI.
We consider  the energy consumption during storing and the diversity of the  data process including sampling, transmitting and receiving, which consume different power levels.
The problem is then mapped as the budget-dynamic Multi-Arm Bandit (MAB) problem by treating the energy as the budget and the data process as arm pulling.
This paper proposes an Opportunistic Duty Cycling (\protocol) scheme to improve the energy efficiency while satisfying the perpetual network operation.
\protocol chooses the proper opportunities to store the harvested energy  or to spend it on the data process based on the historical information of the energy harvesting and the VoI of the processed data.
 With this scheme, each sensor \textbf{node} need only estimate the ambient natural energy in short term so as to reduce the computation and the storage for the historical information. It also can distributively adjust its own duty cycle according to its local historical information.
This paper also conducts the extensive analysis on the performance of our scheme \protocol, and the theoretical results validate the regret, which is the difference between the optimal scheme and ours.
Our experimental results also manifest the promising performance of \protocol.
\end{abstract}
\IEEEpeerreviewmaketitle
\section{Introduction}
\label{section:introduction}
As a promising technique, the great success of Wireless Sensor Networks (WSNs) has been witnessed over a variety of critical applications in recent years~\cite{Mainwaring2002}.
One common constraint,  impeding the wider application of this kind of networks, is the limited energy supply.
To extend the network life or even to support the perpetual network operation, two major techniques have been severally applied to WSNs: energy harvesting~\cite{shen2009solarmote}\cite{tang2011cool}\cite{jiang2005perpetual} and duty cycling~\cite{ghidini2011energy}.
Energy harvesting can supply the sensor node with the extra energy from the ambient energy resources while the duty cycling technique can save energy so as to extend the network lifetime.
But the tiny energy-harvesting module in the solar sensor networks cannot harvest enough energy to support the network with full duty cycle normally~\cite{gu2009esc}\cite{Shen2013EFCon}.
Some existing works combine the energy harvesting and duty cycling techniques to achieve the permanent network operation, \ie, meeting the \emph{energy neutral operation}~\cite{kansal2007power}\cite{moser2010adaptive}.
These existing works estimate the amount of the active time for a period in advance, such as at the initialization phase of the period~\cite{jiang2005perpetual}\cite{moser2010adaptive}, or the average amount of the active time for some periods over a long duration, such as a season~\cite{Buchli2014sensys}.
However, there are several facts ignored by the existing works.

1) Imperfect charge efficiency. In practice, the charge efficiency of the battery for the solar powered sensor node is often less than $75\%$~\cite{ding2000battery}, which means that it indirectly wastes 25\% energy if using the stored energy.
 Another choice, capacitor, suffers high leakage~\cite{zhu2009leakage}. 

2) Variable data importance. In WSNs, the  data redundancy is the common phenomenon. Meanwhile, if introducing the Value of Information (VoI) for the processed data\footnote{In this paper, the data process or to process data means sampling, receiving and transmitting data.}, the more important data has higher VoI. It can obtain higher energy efficiency to process the more important data.

3) Random natural energy. Some natural energy, such as solar or wind energy, is shown to be random~\cite{gu2009esc}\cite{Shen2013EFCon}\cite{ho2010markovian}, so as hard to accurately predict the profiles of the future energy for long term  because of the unpredictable weather and disturbance.

Section \ref{sec:motivation} illustrates some detailed technical evidences and examples to illustrate the above observations. We find that it is still an open problem to improve the efficiency to exploit the ambient energy.

Notice that the energy consumption caused by the imperfect charge efficiency can be decreased if the harvested energy is directly used rather than stored in the battery.
Considering the data importance, the sensor node can arrange right moments to process data and to harvest energy so as to improve the energy efficiency, which is defined as the average VoI obtained per unit energy consumption in this paper.
To do this, we propose the Opportunistic Duty Cycling (\protocol) scheme to catch the features: the dynamic profile of the  energy harvesting, the variable VoI of the data, and the easiness to estimate the harvested energy in short term.
Meanwhile, \protocol considers the diversity of the data process including three actions: data sampling, transmitting and receiving, which consume different power since they have much impact on the energy efficiency.
This paper  then maps the opportunistic duty cycling as the gambling game: Multi-Arm Bandit (MAB)~\cite{Liu2010Distributed}.
In the game, the sensor node is treated as the gambler.
The gambler decides its next action (sampling, receiving, transmitting or storing energy) step by step based on its estimation for the harvested energy and the VoI of the data to process in the subsequent time.

In the real applications of the energy-harvesting WSNs, the data process and energy harvesting are highly dynamic.
Under the MAB game, each sensor node can determine its next state according to its historical information in short term so as to deal with the dynamic feature.
The goal of the gambling game is to maximize the energy efficiency for each sensor node.
Clearly, in order to achieve this goal, each sensor node should carefully decide its next action while adhering to the energy constraint.
Notice that to meet the energy neutral operation and to improve the energy efficiency usually contradict to each other when adjusting the duty cycle.
The former goal requires each sensor node to short its duty cycle while the later requires longer one to obtain the overall VoI as much as possible.
To achieve the bi-criteria object, this paper adjusts  the VoI threshold according to the historical information.

\textbf{Contributions.}
The contributions of this paper include:

1) This paper adjusts the duty cycle by considering the imperfect charge efficiency and the VoI of the data while meeting the energy neutral operation.
We map the new duty cycling problem as the budget-dynamic MAB problem.
To our best knowledge,  this is the first work to formulate and study the problem.

2) This paper designs \protocol scheme to achieve the bi-criteria object.
With \protocol, each sensor node can distributively determine the action to take for the next time slot by running the MAB with  the previous reward and  harvested energy.
An algorithm, called \protocol, is designed to implement the \protocol scheme.
We theoretically analyze the  performance of \protocol by measuring the regret, the difference between the optimal scheme and \protocol.

3) The extensive experiments are also conducted to evaluate the performance of our scheme.
In the experiments, because of the hardness to find the optimal scheme, we propose two baseline approaches: a Centralized and Off-line duty cycling Algorithm (\schemeO), and a Simple Duty Cycling (\schemeS).
\schemeO has the complete knowledge of the natural energy and the data VoI in advance.
\schemeS predicts the energy to harvest and calculates the duty cycle in advance as the algorithm given in the reference~\cite{jiang2005perpetual}.
The experimental results show that the average energy efficiency achieved by our scheme is only 16.02\% lower than that of \schemeO, and 69.09\% higher than that of \schemeS.

\textbf{Road map.} The following context of the paper is organized as follows.
Section \ref{sec:motivation} describes the motivation based on our preliminary experiments, and formulates the opportunistic duty cycling problem in Section~\ref{sec:problem}.
The problem is mapped as the budget-dynamic MAB problem, and \protocol is presented in
Section \ref{sec:online duty_cycling} with its performance analysis in Section \ref{sec:performance analysis}, while the experimental results are discussed in Section \ref{sec:experiment}.
In Section \ref{sec:related}, we review the related works on the energy harvesting module and the duty cycling schemes for WSNs and conclude this paper in Section \ref{sec:conclusion}.
\section{Preliminary Experiments and Motivation}
\label{sec:motivation}
 This work is motivated by the following observations.
Firstly, the inherent hardware property of the energy harvesting module leads to time varying charge efficiency.
In practice, the average charge efficiency of the battery  for the solar powered sensor node is often less than $75\%$~\cite{ding2000battery}.
Secondly,  the random environmental factors, such as the shadow of clouds, can also decrease the charge efficiency.
Thirdly,  the data VoI varies over time and is different among the nodes.
These observations leave the existing duty cycling schemes unsuitable, and motivate us to design the new duty cycling scheme.
\subsection{Dynamic Energy Harvesting and Storage }\label{subsect: dynamic harvested energy}
\begin{figure}\centering
\subfigure[Energy harvested by one sensor node in three days]{\label{subfig:one node in several days}\includegraphics[scale=.8,bb=50 676 351 770]{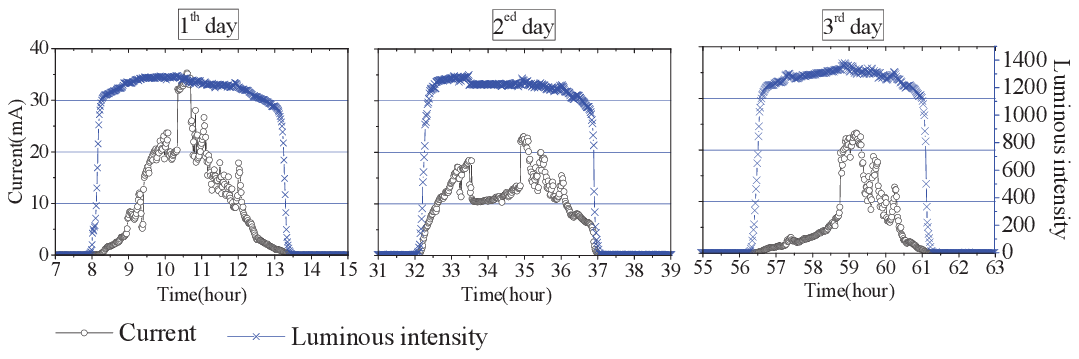}}
\hspace{0.3cm}
\subfigure[Energy harvested by three sensor nodes in one day]{\label{subfig:Several nodes in one day}\includegraphics[scale=.8, bb=35 626 338 724]{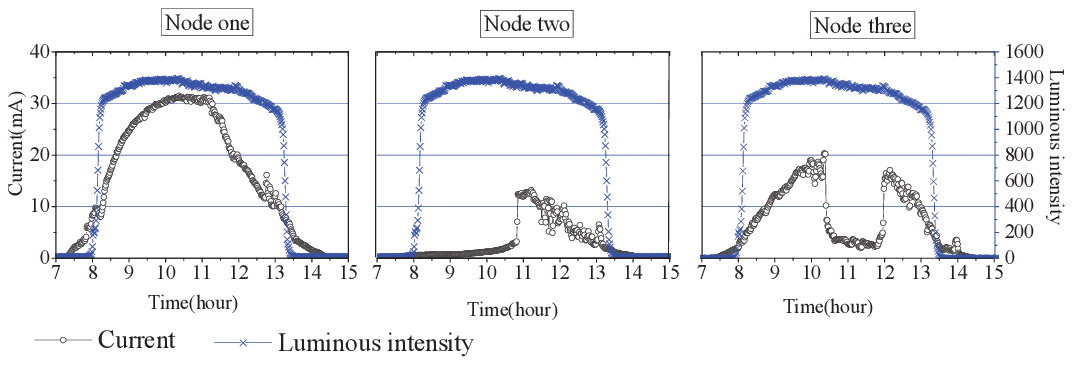}}
\caption{\label{fig:solar experiment} Current indicates the amount of the harvested energy.
(a) Energy profile diverse with time. (b) Different sensor nodes has  different profiles.}
\end{figure}

The unpredictable environmental factors cause the diversity of the energy profiles among the sensor nodes as illustrated in Figure~\ref{fig:solar experiment}.
The experiment results in Figure~\ref{subfig:one node in several days} indicate that the same sensor node usually has different energy profiles in several days even under the similar weather conditions.
More so, the energy profiles for several different sensor nodes vary a lot during one day because of the different locations as shown in Figure~\ref{subfig:Several nodes in one day}.
Similar phenomenon was also observed in previous works~\cite{gu2009esc}\cite{kansal2004performance}\cite{zhu2009leakage}.
Some works model the solar energy harvesting as a first-Markov random process~\cite{ho2010markovian}.

The  time to consume or store harvested energy has great impact on the energy efficiency.
Due to the imperfect charge efficiency, denoted by $\charge$, the relation between the harvested energy $\energy^h$ and the actual stored energy $\energy^s$ is $\energy^s=\charge\energy^h$ for some charge efficiency $\charge<0.75$.
The solar panels in the most existing  solar modules, such as SolarMote~\cite{shen2009solarmote}, Prometheus~\cite{jiang2005perpetual} and AmbiMax~\cite{park2006ambimax}, have the rated current of about 20mA.
Meanwhile, the working current of the sensor node, such as TelosB, is about 20 mA for receiving and about 19 mA or more for transmission.
If the sensor node powers its antenna with the harvested energy (20mA) directly, then the antenna can work normally.
Otherwise, if the sensor node stores the harvested energy with the power 20mA, the actual stored energy is 20$\times0.75$=15mA\footnote{There is a fault voltage to support the normal operation of the sensor node, such as 3 V for the TelosB and MICA nodes.
This paper ignores the voltage for simplicity, and thus represents the power by the unit: mA.} given $\charge=0.75$, which means that 5mA harvested energy is wasted.
The power of the stored energy is thus too low to support the normal  operation of the sensor node.
\subsection{VoI of Data}
\begin{figure}
\begin{minipage}[t]{.23\textwidth}
\centering
\includegraphics[scale=.8, bb=84 500 202 583]{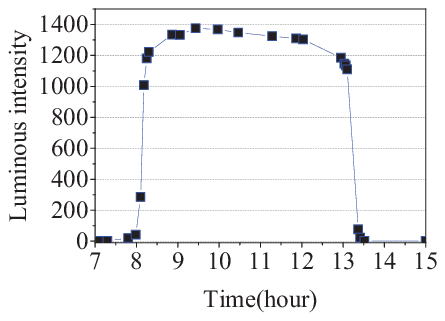}
\caption{\label{subfig:reduced sample rate} Data process is reduced greatly while a little VoI is lost.}
\end{minipage}
\hspace{.3cm}\begin{minipage}[t]{.23\textwidth}
\centering
\includegraphics[scale=0.9,bb=247 386 348 462]{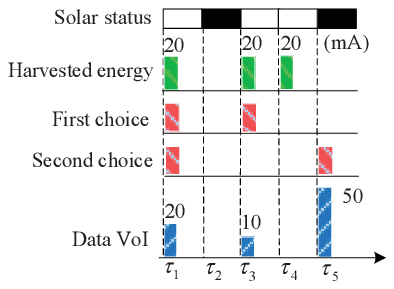}
\caption{\label{subfig:energy allocation affecting information value}Example for different data process choice.}
\end{minipage}
\end{figure}

The  limitation of the harvested energy compels each sensor node to preferentially process  the data with high VoI.
According to Information Theory, the data importance can be indicated by the VoI, denoted by $\reward$~\cite{padhy2010utility}.
The Kullback-Leibler (KL) divergence measure can calculate the VoI by qualifying the difference between two probability distributions: $\prob_1(t)$ and $\prob_2(t)$ as follows.
\begin{equation}\label{equ:KL distance}
\reward_{KL}(\prob_1(t),\prob_2(t))=\int\prob_1(t)\log\frac{\prob_1(t)}{\prob_2(t)}
\end{equation}
With the concept of VoI, the sensor node chooses the important data (\ie with high VoI) to process.
The times to process data then can be decreased so much energy can be saved while the overall VoI is preserved.
For example, when reducing the times to sample the luminous intensity from Figure~\ref{subfig:Several nodes in one day} to Figure~\ref{subfig:reduced sample rate}, about 92\% energy is saved while the overall VoI lose is preserved under 5\%.
\subsection{Call for Online Energy Allocation}
Since both of the data process and energy harvesting are random processes, each sensor node can  make online decision on how to allocate the harvested energy.
The example in Figure~\ref{subfig:energy allocation affecting information value} illustrates the necessity of the online energy allocation to maximize the overall VoI by carefully scheduling the energy consumption.
In this example,  the sensor node $\node_i$ can harvest 20 mA energy at  the time slots marked with ``white" color solar status, and cannot harvest energy at the ``black" time slots.
Suppose that $\node_i$  requires at least 20 mA energy to support its normal operation at each time slot,  and that the charge efficiency $\charge=0.75$.
When time $t$ goes to $\slot_1$, $\node_i$ can use the harvested 20mA energy directly to process the first data with 20 unit VoI.
After $t$ goes to $\slot_3$, $\node_i$ has two choices.
The first choice is that $\node_i$ uses the harvested energy at $\slot_3$ to process the second data, and then obtains 10 unit VoI.
At $\slot_4$, $\node_i$ stores the harvested 20mA energy, and obtains 15mA energy because $\charge=0.75$.
At $\slot_5$, $\node_i$ cannot process data since the stored energy is not sufficient.
The VoI per unit energy that $\node_i$ obtained by the first choice is $\frac{20+10}{20+20}=0.75$.
The second choice is that $\node_i$ stores the 40mA energy harvested at $\slot_3$ and $\slot_4$ and obtains 30mA energy.
It then processes the second data at $\slot_5$, and obtains $50$ unit VoI. The VoI per unit energy that $\node_i$ obtained by the second choice is $\frac{20+50}{20+30}\doteq1.4$.
Obviously, the second choice can result in higher energy efficiency, \ie, the VoI per unit energy, than the first one.
\subsection{Opportunistic Duty Cycling}
\begin{figure}[h]
\hspace{.3cm}\begin{minipage}[t]{.16\textwidth}
\centering
\includegraphics[scale=.9,bb=262 403 334 434]{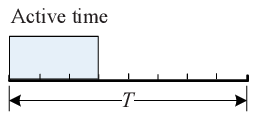}
\caption{\label{subfig:existing duty cycling}Previous duty cycling.}
\end{minipage}
\begin{minipage}[t]{.30\textwidth}
\hspace{.3cm}\subfigure[\label{fig:orpp duty cycling a}Case $a$]{\includegraphics[scale=.9,bb=262 403 334 434]{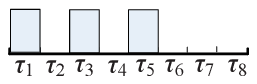}}
\hspace{.2cm}\subfigure[\label{fig:orpp duty cycling b}Case $b$]{\includegraphics[scale=.9,bb=262 403 334 434]{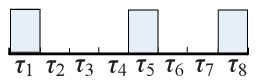}}
\caption{\label{fig:difference between duty cycle and orpportunistic duty cycling} Opportunistic duty cycling. }
\end{minipage}
\end{figure}

From the above facts, we find that the processes of the data process and energy harvesting are highly dynamic.
It can greatly improve the energy efficiency to wake up the sensor node to process data and to hibernate them for storing energy at proper moments.
These facts motivate us to propose the novel opportunistic duty cycling scheme, under which the sensor  nodes can catch the right opportunities to process data or to store  the harvested energy.
Existing works on duty cycling adjust only the duty cycle, \ie, roughly the ratio of the active time to the period as shown in Figure~\ref{subfig:existing duty cycling}.
Under the opportunistic duty cycling, the slots to be active are also considered as the example in Figure~\ref{fig:difference between duty cycle and orpportunistic duty cycling}, where the period composes of 8 slots.
The set of slots to be active may be different as the cases $a$ and $b$ in Figure~\ref{fig:difference between duty cycle and orpportunistic duty cycling} although the duty cycles under both cases are same, \ie, $\frac{3}{8}$.
The reason to adjust the duty cycle in this way is that it may result in different energy efficiency to be active in different slots.
The goal of the opportunistic duty cycle is to adjust the duty cycle and the moments to be active so that the energy efficiency can be improved under the constraint of the energy neutral operation.

Most symbols used in this paper are summarized in Table~\ref{table:notations}.
\begin{table}[h]
\caption{Symbol and meaning\label{table:notations}}
\vspace*{-12pt}
\begin{center}
\def\temptablewidth{0.45\textwidth}
{\rule{\temptablewidth}{1pt}}
\begin{tabular*}{\temptablewidth}{@{\extracolsep{\fill}}c|l|c|lr}\centering
Sym. & Description &Sym. & Description \\ \hline
  $\period$&\text{Period}& $\arm$&\text{Arm of bandit machine}\\
$\node$&\text{Node}&$\pullarmnumtotal$&\text{\# of pulling all arms}\\
$\set$&\text{Set of slots}&$\pullarmnumber$&\text{\# of pulling one arm}  \\
$\armnum$&\text{\# of arms} &$\reward$&\text{Reward/VoI} \\
$\pullvariable$&\text{Pull variable}&$\rewarde$&\text{Estimation of $\reward$}\\
$\route$&\text{Route}&$\rewardth$&\text{Threshold of VoI}\\
$\rewardd$&\text{Reward difference}&$\VoIbound$ &Upper-bound of VoI \\
$\regret$&\text{Regret}&$\prob$&\text{Probability} \\
$\scheme$&\text{Scheme}&$\cost$&\text{Cost/Energy consumption}\\
$\solarstate$&\text{Solar state}&$\energy$&\text{Energy processed in  slot $\slot$}\\
$\charge$&\text{Charge efficiency}&$\Energy$&\text{Energy processed till slot $\slot$}\\
$\expect$&\text{Expectation}&$\remaindata$&\text{VoI of remaining data}\\
$\paraa,\parab,\parac$ &\text{Coefficients} &$\vectorpara,\vectorfeature$&\text{Coefficient vectors} \\
$\parad$&\text{Padding function}& \\
  \end{tabular*}
{\rule{\temptablewidth}{1pt}}
\end{center}
\end{table}
\section{System Model and Problem Formulation}
\label{sec:problem}
 \subsection{Network and Energy Model}\label{subsect:network model}
Given the  network with a sink and some nodes $\node_i$, $i=1,2,\cdots$,   each node  is assumed to have at least one stable route leading to the sink.
A period $\period$ composes of $|\period|$ time slots $\slot_i$, $i=1,\cdots,|\period|$.
Each node  is equipped with a micro-scale energy-harvesting module, and its antenna works under the half-duplex mode.
It cannot receive and transmit data at same time.
It is equipped with one battery to store energy with the initial energy $\energy_0$.
Because of the limited hardware, the battery cannot support the operation of the sensor node when it is being charged by the energy-harvesting \cite{shen2009solarmote}\cite{zhu2009leakage}\cite{park2006ambimax}.
Meanwhile,   the power of the micro-solar panel is also too low to support the normal operation of the sensor node and the battery charging simultaneously in most time as the experimental result in Figure~\ref{fig:solar experiment}.
We thus assume that the limited harvested power cannot support the normal operation of the sensor and antenna simultaneously.

For each sensor node $\node_i$, the different power levels are required to support data sampling, receiving, transmitting and storing the harvested energy, respectively denoted by $\cost^s$, $\cost^r$, $\cost_i^t$ and $\cost_i^g$.
$\cost^s$ and $\cost^r$ are constant and same over all sensor nodes.
The VoI, denoted by $\reward_i(\slot)$, is measured by Equation~(\ref{equ:KL distance}).
Denote the amount of energy harvested by a single sensor node  at time slot $\slot$ by $\energy^h(\slot)$.
The harvested energy $\energy^h(\slot)$, $\slot\in\period$, over a period can be modelled as the first-order stationary Markov process~\cite{ho2010markovian}\cite{ventura2011markov}. The processed data is the same.
Each solar panel can support its   node's normal operation  or can charge its node's battery if and only if its harvested energy is over a threshold $\energyTh$.
Let $\solarstate=1$ if the power of the harvested energy is over the threshold,  and 0 otherwise.
\subsection{Opportunistic Duty Cycling Problem}\label{subsect:problem formulation}
The opportunistic duty cycling can be formalized as the optimization problem.
The goal of \protocol is to maximize the overall VoI collected at the sink as given in Equation~(\ref{equ:max info}), while satisfying the energy neutral operation under the constraint of the energy harvesting randomness  in Equation~(\ref{equ:energy neutral operation condition}).
\begin{equation}\label{equ:max info}
\max\sum\limits_{\slot\in\period}\reward_{sink}(\slot)
\end{equation}
where $\reward_{sink}(\slot)$ denotes the VoI  received by the sink at $\slot$.
At the time slots in the sets $\set^s$, $\set^r$ and $\set^t$,  the sensor node $\node_i$ samples, receives and transmits data respectively.
At the time slots in the set $\set^g$,   $\node_i$ stores the harvested energy into its battery and thus $\solarstate=1$ at every slot in $\set^g$.
To maintain the perpetual operation, the consumed energy should be less than the harvested.
\begin{equation}\label{equ:energy neutral operation condition}
|\set^s|\cost^s+|\set^t|\cost^t_i+|\set^r|\cost^r+|\set^g|\cost^g_i\leq\sum_{\slot\in\period}\energy_i^h(\slot)
\end{equation}

According to the assumption in the  subsection~\ref{subsect:network model}, the antenna is half-duplex so the sets $\set^r$, $\set^t$ has no common element.
Meanwhile, the four sets: $\set^g$, $\set^s$, $\set^r$, and $\set^t$ have no common element because of the limited hardware and harvested energy.
The  four sets thus satisfy the following condition.
\begin{equation} \label{equ:our active set condition}
\left\{ \begin{aligned}
     &\set^g\cup\set^s\cup\set^r\cup\set^t=\period\\
      &\set^r\cap\set^t=\varnothing \text{; and }\set^r\cap\set^s=\varnothing\text{; and }\set^t\cap\set^s=\varnothing\\
      &\set^g\cap\set^s\cup\set^r\cup\set^t=\varnothing
                          \end{aligned} \right.
                          \end{equation}

The core of \protocol scheme is to find these four subsets: $\set^s$, $\set^r$, $\set^t$ and $\set^g$,  so as to solve the optimal problem  in Equation~(\ref{equ:max info}) under the constraint in Equation~(\ref{equ:energy neutral operation condition}) and (\ref{equ:our active set condition}).
\section{Opportunistic Duty Cycling}
\label{sec:online duty_cycling}
This section formulates the opportunistic duty cycling as the budget-dynamic MAB problem~\cite{gittins2011multi}, and then presents our duty cycling scheme: \protocol.
\subsection{Budget-dynamic MAB Problem}\label{subsect:multi arm bandit problem}
\begin{figure}\centering
\subfigure[\label{fig:received and sampled data} Node $\node_i$ has four actions: sampling, receiving, transmitting and storing.]{\includegraphics[scale=.9,bb=233 381 358 455]{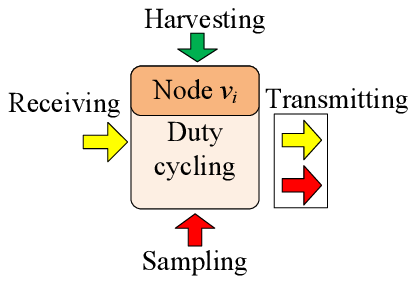}}
\hspace{.5cm}\subfigure[\label{fig:multi armed bandit}Mapping $\node_i$ to a gambler with four arms, $\arm_1$, $\arm_2$, $\arm_3$ and $\arm_4$.]{\includegraphics[scale=.9,bb=244 381 344 455]{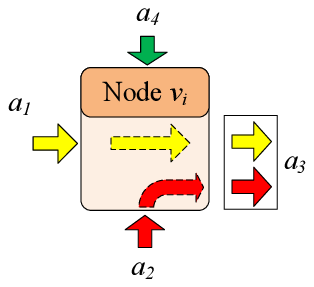}}
\caption{\label{fig:multi arm mapping} Four actions become four arms respectively after mapping a node to a gambler. }
\end{figure}

Let us look into the detailed process of the opportunistic duty cycling in the energy harvested WSNs.
With the harvested energy, each node has two ways to deal: consuming or storing it.
 To store the energy means some energy consumption because of the imperfect charge efficiency, \ie, $\charge<1$.
Otherwise, it spends the harvested energy on the data process.
When no energy to harvest, it must spend the energy in its battery on the data process,  or sleep so as to lose the chance to process data.
Obviously,  each node has to choose one of the four actions: sampling, receiving, transmitting data and storing energy (\ie sleeping), as shown in Figure~\ref{fig:received and sampled data}, by consuming the harvested or stored energy at each time slot.
To maximize the energy efficiency, the node need choose the best action by learning the historical information of the  energy harvesting and data process.
Since the energy harvesting and data process  are the Markov process,  the conditional probability (given the historical information) that the harvested energy and VoI of the data are at certain levels   at the beginning of slot $\slot$  is a sufficient statistic for the design of the optimal actions in the slot $\slot$~\cite{smallwood1973optimal}.
Each node thus need not record the long historical information, and can estimate the  VoI for the next time slot by counting the probability that the power and VoI of the data are at certain levels during the previous time slots in short term.

If treating the sensor node as the gambler, the harvested energy is  the budget of the gambler and the four actions represent the four arms of the bandit machine as shown in Figure~\ref{fig:multi arm mapping}, the opportunistic duty cycling can be formulated  as the budget-dynamic MAB problem.
Pulling the  arms $\arm_1$, $\arm_2$, $\arm_3$ and $\arm_4$ are   the four actions: data receiving, sampling, transmitting and energy storing.
In the MAB problem,  the gambler pulls one of the bandit machine's arms by costing some budget.
The bandit machine then returns the gambler with some reward each time.
For simplicity, we take the VoI of the processed data as the reward.
For example, the node receives a data, whose VoI is $\reward$, and then the reward returned to the node is $\reward$.
The goal of the gambler is to maximize the overall reward under its budget constraint by a series of arm pullings.
In this paper, the harvested energy, \ie the budget, is dynamic, so the  problem in this paper is a new variation of the classical stochastic MAB problem: the budget-dynamic MAB problem.
By mapping the opportunistic duty cycling problem to the MAB problem, the goal to maximize the energy efficiency is equivalent to maximizing the reward given the budget.

Since one sensor node is treated as one gambler in the MAB problem, it means that the solution to the problem is implemented distributively.
The challenge to solve the problem is to prove the distributive scheme can guarantee  the global maximization of the overall VoI.
Recall that the goal is to maximize the overall VoI of the processed data as given in Equation~(\ref{equ:max info}).
Thus, the straightforward idea is to maximize the VoI of the data processed by each node including the data sampling, receiving and transmitting.
The VoI caused by the three actions is denoted by $\reward^s$, $\reward^r$ and $\reward^t$ respectively.
Meanwhile, the overall VoI of the data received by the sink can be maximized only if each node transmits its sampled or received data to the neighbors in the next-hop as much as possible.
In the following context, we consider the more general case than that given in Equation~(\ref{equ:our active set condition}) to state the straightforward idea. Notice that the case in Equation~(\ref{equ:our active set condition}) is covered by the following statements.
Let $\remaindata(\slot)$ denote the overall VoI of the data remaining in $\node_i$'s memory till the end  of  time slot $\slot$.
Recall that  each node cannot receive and transmit data simultaneously  as the constraint in Equation (\ref{equ:our active set condition}).
When the node takes the action to transmit data in $\slot$, there is a balance that  is $\remaindata(\slot)$$=$$\remaindata(\slot-1)$$+$$\reward^s(\slot)$$-$$\reward^t(\slot)$ at time slot $\slot$, where $\reward^s(\slot)$ is the VoI of the sampled data at the slot $\slot$.
We have  the following equation:
\begin{equation}\label{equ:transmission reward function}
\reward^t(\slot)=\remaindata(\slot-1)-\remaindata(\slot)+\reward^s(\slot)
\end{equation}
Similarly, we have the following equation when the node takes the receiving action.
\begin{equation}\label{equ:receiving reward function}
\remaindata(\slot)=\reward^r(\slot-1)+\remaindata(\slot-1)+\reward^s(\slot-1)
\end{equation}
where $\reward^r(\slot-1)$ is  the VoI of the  received data at time slot $\slot-1$ respectively. $\reward^s(\slot-1)$ and $\reward^r(\slot-1)$ may be zero since the action: data transmitting or energy storing, may be taken.
Considering the special case that only one of the four items: $\reward^t(\slot)$,$\reward^s(\slot)$, $\reward^r(\slot)$ and $\reward^s(\slot)$ can be the value over zero, Equation (\ref{equ:transmission reward function}) and (\ref{equ:receiving reward function}) satisfy the constraints in Equation (\ref{equ:our active set condition}).

Recall that each node has at least one routing connecting with the sink as the statement in Section~\ref{subsect:network model}.
Let $\route_k$ denote the set of nodes that are $k$ hops away from the sink,  $k=1,2,\cdots$.
The overall reward of the whole network can be calculated as $\sum\limits_{\slot\in\period}\reward_{sink}(\slot)=\sum\limits_{\slot\in\period}\sum\limits_{\node_i\in\route_1}\reward^t_i(\slot)$ in the period $\period$, where $\reward^t_i(\slot)$ is the VoI of the data transmitted by the node $\node_i$ at the time slot $\slot$.
The following theorem proves that $\sum\limits_{\slot\in\period}\reward_{sink}(\slot)$ can be maximized through maximizing the overall reward of each single node.
This paper decomposes the overall reward of the whole network to that of each node by the following theorem.
\begin{theorem}\label{theorem: equivalence of network total reward}
Assume each node has at least one route connect to the sink, the total reward  of all nodes accumulated over the overall period  equals to the total reward received by the sink over the same period.
\end{theorem}
\begin{IEEEproof}
The intuitive idea of the proof is that all of the data received by the sink must be sent or relayed by the intermediate nodes in the network.
Let $\node_0$ denote the sink, and suppose that the network starts at the time slot $\slot=0$.
When $\slot=0$, \ie, the network does not begin to run, each node $\node_i$ does not receive or sample any data so $\remaindata_i(\slot_i=0)=0$.
In an arbitrary time slot $\slot>0$, the VoI of the data received by the sink is that the relay node $\node_i\in\route_1$ transmits at the same slot.
That is
\begin{equation}\label{equ:proof for theorem one a}
\reward_{sink}(\slot) =\sum\limits_{\node_i\in\route_1}\reward_i^t(\slot) \nonumber 
\end{equation}
Thus, to maximize $\reward_{sink}(\slot) $ is equivalent to maximizing the data traffic of each node away one-hop from the sink when the time slot $\slot$.
According to Equation~(\ref{equ:transmission reward function}),  the right side of the above equation can be rewritten as the following:
\begin{equation}\label{equ:distrution equivalent 1}
 \reward_i^t(\slot)=\remaindata_i(\slot-1)-\remaindata_i(\slot)+\reward_i^s(\slot), \node_i\in\route_1
\end{equation}
Notice that any data sampled or received at time slot $\slot$ can be transmitted after $\slot$.
The transmitted data $\reward_i^t(\slot)$ must come from the remaining data $\remaindata_i(\slot-1)$.
The last two items $\remaindata_i(\slot)$ and $\reward^s(\slot)$ have no contribution to $ \reward_i^t(\slot)$.
Before the time slot $\slot$, $\node_i$ ($\node_i\in\route_1$) must receive or sample the data to record it in $\remaindata_1(\slot^t-1)$.
Otherwise, it has no data to transmit in $\slot$.
The data that the sensor node chooses to transmit at time slot $\slot$ must be received or sampled in some time slot $\slot'$ before $\slot$, \ie, $\slot'<\slot$.
When the sensor node transmits the data in $\slot$, the time $\slot-1$ or $\slot'$ ($\slot'<\slot-1$) at which the data is received or sampled has no affection on the transmission of the data.
For easy to understand the proof, we can assume that the data that the sensor node chooses to transmit at time slot $\slot$ is received or sampled in $\slot-1$.
Meanwhile, the data received by the sensor nodes in the layer $\route_k$ must be transmitted by those in the layer $\route_{k+1}$ so we have the following equation:
\begin{equation}\label{equ:equal receiving and transmitting}
\sum_{\node_i\in\route_k}\reward^r_i(\slot)=\sum_{\node_j\in\route_{k+1}}\reward^t_j(\slot)
\end{equation}
According to Equation (\ref{equ:receiving reward function}) and (\ref{equ:proof for theorem one a}), the VoI of the data received by the sink till time slot $\slot$ is:
\begin{align}\label{equ:distrution equivalent 2}
&\reward_{sink}(\slot)=\sum\limits_{\node_i\in\route_1}\reward_i^t(\slot)=\sum\limits_{\node_i\in\route_1}[\remaindata_i(\slot-1)-\remaindata_i(\slot)+\reward_i^s(\slot)]\nonumber\\
&=\sum\limits_{\node_i\in\route_1}\remaindata_i(\slot-1)-\sum\limits_{\node_i\in\route_1}[\remaindata_i(\slot)+\reward_i^s(\slot)]\nonumber\\
&=\sum\limits_{t=0}^{\slot-1}\sum_{\node_j\in\route_2}\reward_j^t(t)+ \sum\limits_{t=0}^{\slot}\sum\limits_{\node_i\in\route_1}\reward_i^s(t) -\sum\limits_{\node_i\in\route_1}\remaindata_i(\slot)
\end{align}
In the last equality of the above equation, the first item is the sum of the traffic of the sensor nodes in the layer $\route_2$, which contributes to the VoI of the data received by the sink, \ie, $\reward_{sink}(\slot)$  during time slot $\slot-1$.
In other words, the VoI of each sensor node $\node_j\in\route_2$ must be maximized in $\slot-1$ before the overall VoI $\reward_{sink}(\slot)$ can be maximized at time slot $\slot$ since the last two items have no contribution to $\reward_{sink}(\slot)$ in $\slot-1$  according to the statement below Equation~(\ref{equ:distrution equivalent 1}).

Similarly, we can deduce $\reward_{sink}(\slot)$ in Equation (\ref{equ:distrution equivalent 2}) back to the sum of the VoI of the data transmitted by the sensor nodes in the layer $\route_k$ during time slot $\slot-k+1$.
Therefore, the overall VoI of the sink in the period $\period$, \ie $\sum_\slot^\period\reward_{sink}(\slot)$, can be maximized by maximizing the VoI of the data transmitted by each sensor node in each layer over a series of time slot $\slot$, $\slot\in\period$.
\end{IEEEproof}
\subsection{\protocol}\label{subsect:multi arm bandit game for duty cycling}
This block  presents the detailed design of our scheme:  \protocol.
In order to achieve the energy neutral operation, a parameter, called VoI threshold $\rewardth$, is introduced  to control the amount of energy that each sensor node can consume in each time slot.
Because of the randomness of the harvest energy, $\rewardth$ should be updated continuously.
The Adaptive VoI Adjustment  (\alg) algorithm is designed to  update the threshold $\rewardth$.

\subsubsection{\protocol algorithm}
Recall that the goal of \protocol is to maximize the VoI of each sensor node, \ie to solve the budget-dynamic MAB problem, so that the overall VoI can be maximized according to Theorem \ref{theorem: equivalence of network total reward}.
Imagine that taking an action corresponds to placing an item into the knapsack.
The expected reward by taking the action equals to the item's value and the energy consumption for the action is the item's weight.
The total harvested energy till $\slot$ is then the weight capacity of the knapsack at $\slot$.
Therefore, the budget-dynamic MAB can be reduced to the unbounded knapsack problem at each time slot $\slot$.
We borrow the idea of the density-ordered greedy algorithm~\cite{kohli2004average} to solve the problem.

During solving the budget-dynamic MAB problem by the density-ordered greedy algorithm, the key step is to estimate the VoI that each action will obtain at the next time slot $\slot$, so that the sensor node $\node_i$ can take those actions with the highest   energy efficiency.
Auer introduced the  Upper Confidence Bound  (UCB) to calculate the estimated VoI of each action~\cite{auer2002finite}.
The most popular UCB, called UCB-1, relies on the upper-bound VoI $\rewarde'_j(\slot)+\parad'_j(\slot)$ obtained by taking the action $\arm_j$, where  $\parad'_j(\slot)$ is a padding function.
A standard expression of the function is $\parad'_j(\slot)=\VoIbound\sqrt{\frac{\parae'\ln\pullarmnumtotal(\slot)}{\pullarmnumber_j(\slot)}}$, where $\VoIbound$ is the upper-bound on the reward/VoI,  $\parae'>0$ is some appropriate constant, $\pullarmnumber_j(\slot)$ is the number of taking action  $\arm_j$  till $\slot$, $\pullarmnumtotal(\slot)$ is the overall number of actions that the sensor node $\node_i$ has taken till $\slot$, and $\rewarde'_j(\slot)$ is the estimation of the action $\arm_j$'s expected reward for the slot $\slot$ at the end of the slot $\slot-1$.
In order to improve the energy efficiency, the  upper-bound VoI per unit cost can be calculated as $\rewarde_j(\slot)+\parad_j(\slot)=(\rewarde'_j(\slot)+\parad'_j(\slot))/\cost_j$ by taking the cost $\cost_j$ into consideration.
We have $\rewarde_j(\slot)=\rewarde'_j(\slot)/\cost_j$ and $\parad_j(\slot)=\VoIbound\sqrt{\frac{\parae_j\ln\pullarmnumtotal(\slot)}{\pullarmnumber_j(\slot)}}$, where $\parae_j=\parae'/\cost^2_j$.
Notice that the remaining energy $\Energy(\slot)$  till time slot $\slot$ composes of the energy remained in its battery $\Energy(\slot)$ and possibly harvested energy at $\slot$,  \ie, $\Energy(\slot)=\Energy(\slot-1)+\solarstate(\slot)\energy^h(\slot)$.
Thus, the unbounded knapsack problem can be formulated as the following problem with the time-dependent energy bound $\Energy(\slot)$.
\begin{align}\label{equ: density ordered knapsack}
&\max\sum\limits_{j=1}^\armnum \pullvariable_j(\slot)(\rewarde_j(\slot)+\parae_j)\\
&s.t.\ \sum\limits_{j=1}^\armnum \pullvariable_j(\slot)\cost_j\leq\Energy(\slot),
 \forall j,\slot: \pullvariable_j(\slot)\in\{0,1\} \label{equ;constraint of desnsity of ordered knapsack}
\end{align}
where $\pullvariable_j(\slot)$ is a bool indicator. $\pullvariable_j(\slot)=1$ if the action $\arm_j$ is taken at $\slot$, and otherwise $\pullvariable_j(\slot)=0$.
$\cost_j$ is the energy consumption to pull the arm $\arm_j$ once.
The constraint in Equation~(\ref{equ;constraint of desnsity of ordered knapsack}) means that the energy consumption at time slot $\slot$ is constrained by $\Energy(\slot)$.
$\rewarde_j(\slot)$ can be calculated as the average reward received by pulling arm $\arm_j$ till $\slot-1$.
\begin{equation}\label{equ:information estimation}
\rewarde_j(\slot)=\sum\limits_{t=1}^{\slot-1}\frac{\pullvariable_j(t)\reward_j(t)}{\cost_j\pullarmnumber_j(\slot-1)}
\end{equation}
The problem defined in Equation~(\ref{equ: density ordered knapsack}) is NP-hard so this paper uses the density-ordered greedy method~\cite{kohli2004average} to find a near-optimal selection of the sets $\set^s$, $\set^t$ and $\set^r$, \ie to find the integer $\pullvariable_j(\slot)$ so that Equation~(\ref{equ: density ordered knapsack}) is maximized (see step~\ref{algline1} in Algorithm~\ref{alg:mab scheme}).

The capacity of the memory is limited.
Each sensor node thus should keep balance between its output: the transmitted data and its input: the received and sampled data in the long term.
In other words, the times to take the action: the data transmitting, \ie pulling the arm $\arm_3$, is expected to equal to the sum of the times to take the actions: the data sampling and receiving, \ie pulling the arms $\arm_1$ and $\arm_2$.
To do this, we assign each action with some probability.
Let $\pullvariable^*_j(\slot)$ be the solution to the problem in Equation~(\ref{equ: density ordered knapsack}) by the density-ordered greedy method at the time slot $\slot$.
\protocol takes the next action $\arm(\slot)$ with some probability, which is determined by the following equation (see step~\ref{algline2} in Algorithm~\ref{alg:mab scheme}).
\begin{eqnarray}\label{equ:prob to pull arm}
\prob(\arm(\slot)=\arm_j)=
\left\{ \begin{aligned}
\pullvariable^*_j(\slot)/\sum\limits_{j=1}^\armnum\pullvariable^*_j(\slot),\hspace{0.5cm} j=1,2\\
2\pullvariable^*_j(\slot)/\sum\limits_{j=1}^\armnum\pullvariable^*_j(\slot),\hspace{0.7cm}j=3
                          \end{aligned} \right.
\end{eqnarray}
where $\armnum$ is the number of the arms of the bandit machine.
Notice that the arm with the higher upper bound VoI will have higher probability in Equation~(\ref{equ:prob to pull arm}) since the times that it is pulled is higher than others.
\protocol is presented in Algorithm~\ref{alg:mab scheme}, and its performance will be theoretically analyzed on its regret bound in the next section.
In this algorithm, $\cost(\slot)$ is the energy consumed at time slot $\slot$.
For example, if the arm $\arm_j$ is pulled and the consumed energy is $\cost_j$ in $\slot$, then $\cost(\slot)=\cost_j$.
\begin{algorithm}[hptb]
\caption{The \protocol Algorithm}
\label{alg:mab scheme}
\textbf{Input:} $\cost(1)=0$ and $\rewardth(1)=\rewarde(1)=0$;\\
\qquad \textbf{Output:} A sequence of actions;\\
\begin{algorithmic}[1]
\STATE Initialize: $\slot=0$ and $\Energy(\slot)=\energy_0$;
\WHILE{$\slot+=1$, and $\slot\leq|\period|$}\label{alg:begin}
\STATE Update the remaining energy $\Energy(\slot)$ till $\slot$;
\STATE \label{algstep: a1 }Input $\energy^h(\slot)$ and $\cost(\slot)$ into Algorithm~\ref{alg:information threshold scheme} to update $\rewardth(\slot+1)$;
\IF{$\rewarde(\slot)<\rewardth(\slot)$}\label{alg:step store energy}
\STATE Pull arm $\arm_4$ to store energy;
\STATE $\Energy(\slot)=\Energy(\slot-1)+\charge\solarstate(\slot)\energy^h(\slot)$, and go to the step \ref{alg:begin};
\ENDIF
\IF{$\slot\leq \armnum$}\label{algline:loop}
\STATE Initial phase: pull the arms $\arm_i$, $i=1,2,3$ one by one;
\ELSE
\STATE \label{algline1} Calculate $\pullvariable^*_j(\slot)$ by solving the knapsack problem in Equation~(\ref{equ: density ordered knapsack});
\STATE \label{algline2} Take the action $\arm_j(\slot)$ with the highest probability $\prob(\arm(\slot)=\arm_j)$ given in Equation~(\ref{equ:prob to pull arm});
\STATE $\Energy(\slot)=\Energy(\slot-1)+(\solarstate(\slot)-1)\cost_j$;
\ENDIF
\STATE Update the upper bound VoI $\VoIbound_j$ of the action $\arm_j(\slot)$;
\STATE Update $\rewarde(\slot+1)=\max\limits_{\arm_j:j=1,\cdots,\armnum}\rewarde_j(\slot+1)$ by Equation~(\ref{equ:information estimation})\label{alg: step update voi};
\ENDWHILE
\end{algorithmic}
\end{algorithm}
\subsubsection{\alg}
The intuitive idea behind \alg is that each sensor node dynamically estimates  the VoI threshold for the next time slot according to the harvested energy and the consumed energy  in the previous time slots.
The energy neutral operation condition requires each sensor node to consume energy less than the remaining one, \ie $\Energy^h(\slot)\geq\Energy^c(\slot)$, while the sensor node $\node_i$ has to consume energy as much as possible to maximize the total reward in the period.
The best choice is to keep the balance between the remaining and consumed energy in the period, \ie $\Energy(\period)=\Energy^c(\period)$.
We define the following function as the metric to find the balance point.
\begin{equation}\label{equ:metric function}
\lim\limits_{\period\rightarrow\infty}\frac{1}{|\period|}\sum\limits_{\slot=1}^\period[\Energy(\slot)-\cost(\slot)]^2
\end{equation}
Denote  the VoI threshold updated at $\slot$ by $\rewardth(\slot)$.
A proper $\rewardth(\slot)$ ensures that  the sensor node can minimize the average squared deviation of the harvested energy from the consumed energy by Equation~(\ref{equ:metric function}).
To find the proper $\rewardth$, we adopt the adaptive control theory in Algorithm \alg, transforming the threshold determining
    problem as the linear-quadratic tracking problem.
More formally, this paper argues that a first order, discrete-time, linear dynamical system with colored noise for the problem.
This system can be described by the following equation:
\begin{equation}\label{equ:linear systemquadtatic system}
\cost(\slot+1)=\paraa\cost(\slot)+\parab\rewardth(\slot)+\parac\omega_\slot+\omega_{\slot+1}
\end{equation}
In this system, $\cost(\slot+1)$ is refer  to the output of the system, $\rewardth$ is the control, $\omega$ is mean zero input noise, $\paraa,\parab,\parac$ are real-valued coefficients.
The optimal output of the system is to  keep the metric in Equation~(\ref{equ:metric function}) as small as possible in the  period $\period$.
The optimal control law to minimize the tracking error is~\cite{kumar1986stochastic}:
\begin{equation}\label{equ:optimal control law}
\rewardth(\slot)=[\energy^h(\slot)-(\paraa+\parab)\cost(\slot)+\parac\energy^h(\slot)]/\parab
\end{equation}
The coefficients $\paraa$, $\parab$ and $\parac$ are not known in advance, and can be estimated online in our problem by using the standard gradient descent techniques~\cite{kumar1986stochastic}.
Firstly, we define a parameter vector $\vectorpara_\slot\triangleq(\paraa+\parac,\parab,\parac)^T$, and a feature vector $\vectorfeature_\slot\triangleq(\cost(\slot),\rewardth(\slot),-\energy^h(\slot))^T$.
By the two vectors, the optimal  control law in Equation~(\ref{equ:optimal control law}) can be expressed as $\vectorfeature_\slot^T\vectorpara=\energy^h(\slot)$.
The estimated parameter vector $\hat{\vectorpara}$ for $\vectorpara$ then can be defined by the gradient descent update rule as given  by
\begin{equation}\label{equ:gradient descent update}
\hat{\vectorpara}_{\slot+1}=\hat{\vectorpara}_\slot+\mu\vectorfeature_\slot(\cost_{\slot+1}-\vectorfeature^T_\slot\hat{\vectorpara}_\slot)/(\vectorfeature^T_\slot\vectorfeature_\slot)
\end{equation}
where $\mu$ is a positive constant step-size parameter.

Because  each sensor node need store its harvested energy in its battery, the initial energy level $\energy_0$ would better be about half of its full capacity.
The choice of the $\hat{\vectorpara}_
\slot$'s initial value $\hat{\vectorpara}_0$ greatly affects the converge speed of the parameter estimation in Equation~(\ref{equ:gradient descent update}).
$\hat{\vectorpara}_0$ can be set preciously according to preliminary experimental results.
Examining the system in Equation~(\ref{equ:linear systemquadtatic system}), the increment of the control $\rewardth$ results in less data being received or sampled, so less energy consumption. $b$ should be negative.
Set $\vectorfeature_0=(\cost_0,\rewardth(0),-\energy^h(\slot))$.
\begin{algorithm}[hptb]
\caption{\alg}
\label{alg:information threshold scheme}
\textbf{Input:} The harvested energy $\energy^h(\slot)$ and the consumed energy  $\cost(\slot)$ of  $\node_i$ till $\slot$. $\slot=0$.\\
\qquad \textbf{Output:} The updated threshold $\rewardth(\slot+1)$\\
\begin{algorithmic}[1]
\IF{$\slot=0$}
\STATE $\hat{\vectorpara}_\slot=\hat{\vectorpara}_0$ and set $\vectorfeature_0$;
\ENDIF
\STATE Update the parameter vector $\hat{\vectorpara}_{\slot+1}$ by Equation~(\ref{equ:gradient descent update});
\STATE Update the feature and parameter vectors $\vectorfeature_\slot$, $\vectorpara_\slot$;
\STATE Output $\rewardth(\slot+1)$ using Equation~(\ref{equ:optimal control law});
\end{algorithmic}
\end{algorithm}

Considering a special case in which each sensor node can harvest enough solar energy.
Thus, the harvested energy can support each sensor node to operate at each time slot.
However,  each sensor  node cannot harvest sufficient energy usually so $\rewardth(\slot)$ prevents each sensor node from working at every time slot, $\ie$ by reserving some energy at some time slots.
So the harvested energy is stored and will not be consumed completely at every time slot, \ie, $\Energy(\slot)\geq 0$.
\subsection{Common Activity}
\begin{figure}[h]\centering
\begin{minipage}[b]{.22\textwidth}
  \includegraphics[scale=0.57,bb=116 633 301 772]{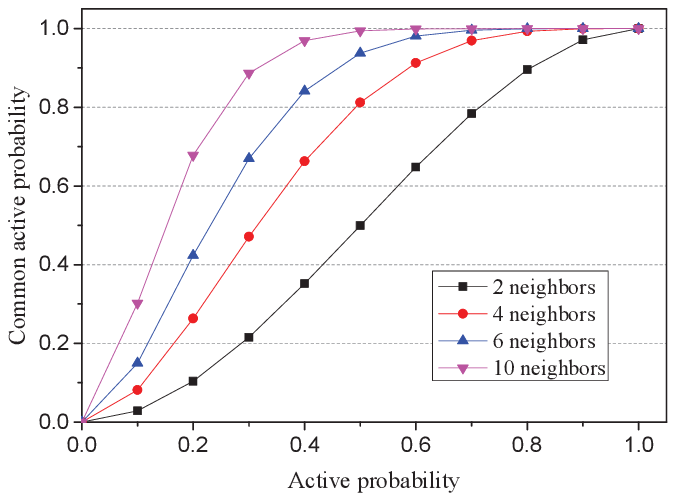}
  \caption{\label{fig:theoretical common active time}Theoretical common active time.}
\end{minipage}
\hspace{0.2cm}\begin{minipage}[b]{.22\textwidth}
  \includegraphics[scale=0.57,bb=111 633 322 778]{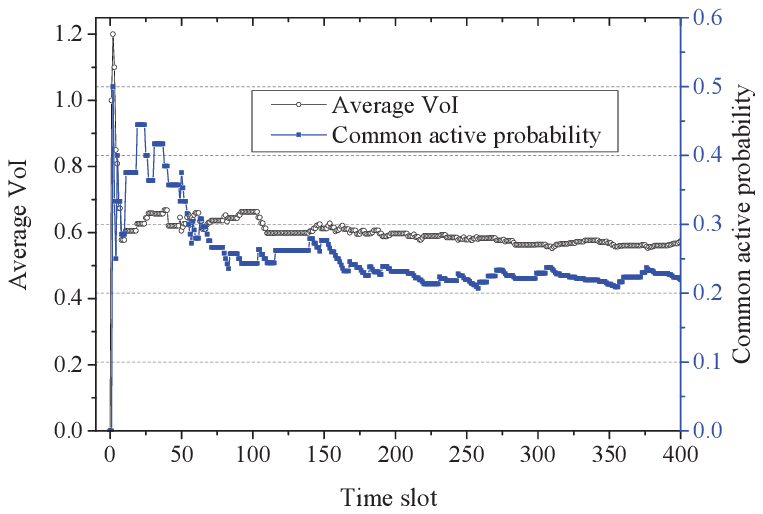}
  \caption{\label{fig:experimental common active time}Experimental common active time.}
\end{minipage}
\end{figure}

A concerned issue is how about the common active time among neighboring nodes under \protocol, which is implemented in the distributive mode.
By Algorithm~\ref{alg:mab scheme},   each node chooses the transmitting and receiving arms with some probability and thus each node has common active, \ie simultaneous waking up,  with some probability in each time slot.
This section shows the probability that one node has common active time with its neighbor theoretically and experimentally.
If the node can communicate with at least one of its neighbors, we say that it has common active time with its neighbor.
Figure~\ref{fig:theoretical common active time} illustrates the theoretical probability that the neighboring nodes have common active time.
When each node has some probability to wake up, \ie active probability,  the common active probability can be easily computed as the y-coordinate.
More neighbors the node has or higher probability it wakes up, it has higher probability to communicate with its neighbor in Figure~\ref{fig:theoretical common active time}.
Figure~\ref{fig:experimental common active time} illustrates the experimental results when one node has two neighbors.
The experimental setting is given in Section~\ref{sec:experiment}.
In the experiment, the common active probability tends to 0.22, and the average data VoI obtained by each action tends to about 0.57.
In each time slot, the node can guarantee a certain probability to communicate with its neighbors. The probability is not quite high but the obtained VoI is not low since the node catches the most important time to communicate.
Next section analyzes that VoI difference of the data processed by the optimal solution and our scheme \protocol.
\section{Performance Analysis}
\label{sec:performance analysis}
This section analyzes the theoretical performance of \protocol by the metric: regret.
Let $\reward_\scheme(\Energy)$ be the total VoI returned by a given algorithm $\scheme$ under the constraint of the variable harvested energy $\Energy$ over a fixed period $\period$.
The expectation of $\reward_\scheme(\Energy)$ is denoted by $\expect[\reward_\scheme(\Energy)]$.
This paper always sticks a superscript ``*" to any instance that is the optimum.
Suppose that $\scheme^*$ is the optimal algorithm for our problem, \ie
\begin{equation}\label{equ:maximize}
\scheme^*=\arg\max\limits_\scheme\expect[\reward_\scheme(\Energy)]
\end{equation}
Thus,  the regret $\regret_\scheme(\Energy)$ of the algorithm $\scheme$ can be formally defined as~\cite{auer2002finite}:
\begin{equation}\label{equ:regret definition}
\regret_\scheme(\Energy)=\expect[\reward_{\scheme^*}(\Energy)]-\expect[\reward_\scheme(\Energy)]
\end{equation}
where is represented by the expectation of the arm with the maximal reward, \ie $\expect[\reward_{\scheme^*}(\Energy)]=\max_{\arm_j=1,2,3}\expect[\reward_j(\Energy)]$, because of the hardness to find the optimal scheme.

In the following context, we analyze the regret of our scheme $\protocol$.
The Hoeffding inequality will be applied in the following analysis, and stated as below:

 The Hoeffding inequality---Let $x_1,\cdots,x_n$ be random variables with common range [0,1] and such that $\expect[x_t|x_1,\cdots,x_{t-1}] =\mu$. Let $S_n =\frac{1}{n}(x_1+\cdots+x_n)$. Then for the constant $a>0$, the probability $\prob(S_n\geq\mu+a)\leq e^{-2na^2}$ and $\prob(S_n\leq\mu-a)\leq e^{-2na^2}$.

Recall that the power of the harvested energy must be higher than the threshold $\energyTh$, and then it can support the normal operation of the sensor node.
Denote  by $\period'$ the time slot set  in which the harvested energy is higher than the threshold $\energyTh$.
$\period'$ is determined by the energy harvesting process, and its expectation can be determined easily if its state transition probability   is previously known.
By the algorithm~\ref{alg:information threshold scheme}, the VoI threshold is continuously adjusted so the sensor node may choose to sleep (\ie to store the harvested energy) in some slots when the harvested energy is higher than the threshold $\energyTh$.
Because of the charging efficiency $\charge<1$, the amount of  the time slots, denoted by $|\period_a|$, in which the harvested energy can support the normal operation of the sensor node under  the algorithm~\ref{alg:information threshold scheme} must be not higher than $|\period'|$.
Thus, we have $|\period_a|\leq|\period'|\leq|\period|$.

Firstly, we analyze the expected times that the arm $\arm_j$, $j=1,2 \text{ or } 3$ is pulled.  The arm $\arm_4$ (storing energy) is not included since it does not return any reward. This is given in the following lemma.
We prove  the following lemma based on the idea of the reference~\cite{auer2002finite} and consider the cost of each arm $\cost_j$, $j=1,2,3$.
\begin{lemma}\label{lemma:epected number of single arm}
For an arbitrary arm $\arm_j$, $j=1,2 \text{ or } 3$, the expected times that it is pulled in the period $\period$,  is:
\begin{equation}
\expect[\pullarmnumber_j(\period)]<(\frac{\cost_{max}}{\cost_{min}})^2\frac{\parae'\ln|\period'|}{\rewardd_j^2}+2
\end{equation}
where $\rewardd_j$ is the difference of the expected reward between the optimal algorithm $\scheme^*$ and the arm $\arm_j$.
$\cost_{max}=\max_{j=1,2,3} \cost_j$ an $\cost_{min}=\min_{j=1,2,3} \cost_j$.
\end{lemma}
\begin{IEEEproof}
Recall that  the step \ref{algline:loop} of the algorithm~\ref{alg:mab scheme} indicates that each arm $\arm_j$, $j=1,2,3$, is pulled once in the first $\armnum$ slots.
Thus, the times to pull $\arm_j$ is $\pullarmnumber_j(\period_a)=1+\sum\limits^{\period_a}_{\slot=\armnum+1}\pullvariable_j(\slot)$, where $\slot\in\period_a$.
Since the algorithm~\ref{alg:mab scheme} is a greedy algorithm, the selected arm has the higher upper-band VoI per unit cost over other arms including the optimal one in each slot $\slot\in\period_a$. So we have the following condition: $\rewarde_j(\slot)+\parad_j(\slot)\geq\rewarde^*(\slot)+\parad^*(\slot)$, \ie $(\rewarde'_j(\slot)+\parad'_j(\slot))/\cost_j\geq(\rewarde^{*'}(\slot)+\parad^{*'}(\slot))/\cost^*$.
In order to satisfy the condition with high probability, at least one of the following inequalities must be satisfied.
\begin{align}
&\rewarde^*(\slot)+\parad^*(\slot)\leq u^*\label{equ:inquality one}\\
&\rewarde_j(\slot)+\parad_j(\slot)\geq  u_j\label{equ:inquality two}\\
& u^{*'}/\cost^*<(u'_j+\parad'_j(\slot))/\cost_j\label{equ:inquality three}
\end{align}
where $u^{*'}$ and $u'_j$ are the reward expectation of the optimal algorithm and the arm $\arm_j$ by our algorithm, which is unknown to the sensor node. $u^*=u^{*'}/\cost^*$ and $u_j=u'_j/\cost_j$.
By using the Hoeffding inequality, the probability that the inequalities in Equation~(\ref{equ:inquality one}) and (\ref{equ:inquality two}) are satisfied is given as follows:
\begin{eqnarray}\label{equ:condition one}
  \prob(\rewarde^*(\slot)\leq u^*-\parad^*(\slot))\leq e^{-4\ln \pullarmnumtotal(\slot)}=\pullarmnumtotal(\slot)^{-4}\nonumber\\
    \prob(\rewarde_j(\slot)\geq  u_j-\parad_j(\slot))\leq e^{-4\ln \pullarmnumtotal(\slot)}=\pullarmnumtotal(\slot)^{-4}
\end{eqnarray}

Recall that $\cost_j>0$ and $u^{*'},u'_j\geq 0$, and then the inequality in Equation~(\ref{equ:inquality three}) implies:
\begin{eqnarray}\label{equ:condition two}
\cost_j u^{*'}<\cost^*(u'_j+\parad'_j(\slot))\Rightarrow\pullarmnumber_j(\slot)< \frac{\cost^{*2}\parae'\ln\pullarmnumtotal(\slot)}{(\cost_ju^{*'}-\cost^*u'_j)^2}&\nonumber\\ \Rightarrow\pullarmnumber_j(\slot)<
\left\{ \begin{aligned}
&\frac{\parae'\ln\pullarmnumtotal(\slot)}{\rewardd_j^2}\hspace{1cm}\cost_j\geq\cost^* \\
&\frac{\cost^{*2}\parae'\ln\pullarmnumtotal(\slot)}{(\cost_j\rewardd_j)^2}\hspace{0.5cm}\cost_j<\cost^*
                          \end{aligned} \right.
\end{eqnarray}
where $\rewardd_j=u^{*'}-u'_j$.

By Equation~(\ref{equ:condition one}) and (\ref{equ:condition two}), the expectation of the times to pull the arm $\arm_j$ thus can be given as follows:
\begin{align}
&\expect[\pullarmnumber_j(\period)]=1+\sum\limits^{\period_a}_{\slot=\armnum+1}\pullvariable_j(\slot)\nonumber\\
&<1+ \max\{\frac{\parae'\ln\pullarmnumtotal(\period_a)}{\rewardd_j^2},\frac{\cost^{*2}\parae'\ln\pullarmnumtotal(\period_a)}{(\cost_j\rewardd_j)^2}\}\nonumber\\
&+\sum\limits^{\period_a}_{\slot=\armnum+1}\{  \prob(\rewarde^*(\slot)\leq u^*-\parad^*(\slot))+\prob(\rewarde_j(\slot)\geq  u_j+\parad_j(\slot))\}\nonumber\\
&\leq1+\frac{\cost^{*2}\parae'\ln\pullarmnumtotal(\period_a)}{(\cost_j\rewardd_j)^2}+\sum\limits^{\period_a}_{\slot=\armnum+1}2\nonumber\pullarmnumtotal(\slot)^{-4}\nonumber\\
&\leq(\frac{\cost_{max}}{\cost_{min}})^2\frac{\parae'\ln\pullarmnumtotal(|\period_a|)}{\rewardd_j^2}+2 \hspace{1cm}(\text{as } |\period_a\rightarrow|\infty)\nonumber\\
&\leq(\frac{\cost_{max}}{\cost_{min}})^2\frac{\parae'\ln|\period'|}{\rewardd_j^2}+2   \hspace{1cm}(|\period_a|\leq|\period'|)
\end{align}
where $\cost_{max}=\max\limits_{\arm_j:j=1,2,3}\cost_j$ and $\cost_{min}=\min\limits_{\arm_j:j=1,2,3}\cost_j$ and $\parae'\geq 1$.
Notice that $\pullarmnumtotal(\period')$ is the total number of times to pull all arms and only one arm can be pulled in each time slot so $\pullarmnumtotal(\period')=|\period'|$.
\end{IEEEproof}

Similarly, we can obtain that the expected times to pull the arm $\arm_4$.
\begin{lemma}\label{lemma:epected number of single arm 4}
The expected times to pull the arm $\arm_4$ in the period $\period$,  is:
\begin{equation}
\expect[\pullarmnumber_4(\period)]<(\frac{\cost_{max}}{\cost_{min}})^2\frac{\parae'\ln|\period'|}{\rewardd_4^2}+1
\end{equation}
where $\rewardd_4=\min\limits_{j=1,2,3}\rewardd_j$.
\end{lemma}
\begin{IEEEproof}
According to   the step \ref{algline:loop} of Algorithm~\ref{alg:mab scheme}, the arm $\arm_4$ will be pulled when $\rewarde(\slot)<\rewardth(\slot)$, which means that at least one of the following inequalities must be satisfied with high probability.
\begin{align}
&\rewarde_j(\slot)+\parad_j(\slot)<  u_j,\hspace{1.4cm}\forall j=1,2,3\label{equ:inquation four}\\
& u^{*'}/\cost^*>(u'_j-\parad'_j(\slot))/\cost_j,\hspace{0.5cm}\forall j=1,2,3\label{equ:inquation five}
\end{align}
By using the Hoeffding inequality, the probability that the inequality  in Equation~(\ref{equ:inquation four}) is satisfied is given as follows:
\begin{eqnarray}\label{equ:condition one for arm 4}
    \prob(\rewarde_j(\slot)<u_j-\parad_j(\slot))< e^{-4\ln \pullarmnumtotal(\slot)}=\pullarmnumtotal(\slot)^{-4}
\end{eqnarray}
The inequality in Equation~(\ref{equ:inquation five}) implies:
\begin{eqnarray}\label{equ:condition two for arm 4}
\cost_j u^{*'}>\cost^*(u'_j-\parad'_j(\slot))\Rightarrow\pullarmnumber_4(\slot)< \frac{\cost^{*2}\parae'\ln\pullarmnumtotal(\slot)}{(\cost_ju^{*'}-\cost^*u'_j)^2}&\nonumber\\ \Rightarrow\pullarmnumber_4(\slot)<
\left\{ \begin{aligned}
&\frac{\parae'\ln\pullarmnumtotal(\slot)}{\rewardd_j^2}\hspace{1cm}\cost_j\geq\cost^* \\
&\frac{\cost^{*2}\parae'\ln\pullarmnumtotal(\slot)}{(\cost_4\rewardd_j)^2}\hspace{0.5cm}\cost_j<\cost^*
                          \end{aligned} \right.
\end{eqnarray}
According to the step~\ref{alg: step update voi} in Algorithm~\ref{alg:mab scheme}, the conditions given in Equation~(\ref{equ:inquation four}) and (\ref{equ:inquation five}) should be satisfied for all arms $\arm_j$, $j=1,2,3$ simultaneously. Therefore, by Equation~(\ref{equ:condition one for arm 4}) and (\ref{equ:condition two for arm 4}), the expectation of the times to pull the arm $\arm_4$ thus can be given as follows:
\begin{align}
&\expect[\pullarmnumber_4(\period)]=\sum\limits^{\period_a}_{\slot=\armnum+1}\pullvariable_4(\slot)\nonumber\\
&<\max_{j=1,2,3}\max\{\frac{\parae'\ln\pullarmnumtotal(\period_a)}{\rewardd_j^2},\frac{\cost^{*2}\parae'\ln\pullarmnumtotal(\period_a)}{(\cost_j\rewardd_j)^2}\}\nonumber\\
&+\sum\limits^{\period_a}_{\slot=\armnum+1}\prod_{j=1}^3\prob(\rewarde_j(\slot)<  u_j+\parad_j(\slot))\nonumber\\
&\leq\max_{j=1,2,3}\frac{\cost^{*2}\parae'\ln\pullarmnumtotal(\period_a)}{(\cost_j\rewardd_j)^2}+\sum\limits^{\period_a}_{\slot=\armnum+1}\nonumber\pullarmnumtotal(\slot)^{-16}\nonumber\\
&\leq(\frac{\cost_{max}}{\cost_{min}})^2\frac{\parae'\ln\pullarmnumtotal(|\period_a|)}{\rewardd_4^2}+1 \hspace{1cm}(\text{as } |\period_a\rightarrow|\infty)\nonumber\\
&\leq(\frac{\cost_{max}}{\cost_{min}})^2\frac{\parae'\ln|\period'|}{\rewardd_4^2}+1 \hspace{1cm}(|\period_a|\leq|\period'|)
\end{align}
where $\arm_4=\min_{j=1,2,3}\rewardd_j$, and $\parae'\geq 1$.
\end{IEEEproof}

Recall that the harvested energy can support the normal operation of the sensor node in at most  $|\period'|$ slots, $\period'\subseteq\period$.
By the lemma~\ref{lemma:epected number of single arm} and \ref{lemma:epected number of single arm 4}, we can analyze the reward regret of Algorithm~\ref{alg:mab scheme}.
\begin{theorem}
For the dynamic energy budget $\Energy(\period)>0$, the expectation of  \protocol's regret  is at  most:
\begin{equation}\label{equ:regret boundary}\sum^{\armnum-1}\limits_{j=1}[(\frac{\cost_{max}}{\cost_{min}})^2\frac{\parae'\ln|\period'|}{\rewardd_j}+2\rewardd_j]+u^{*'}[(\frac{\cost_{max}}{\cost_{min}})^2\frac{\parae'\ln\pullarmnumtotal(|\period'|)}{\rewardd_4^2}+1]
\end{equation}
where $\parae'\geq 1$ is a constant, and $u^{*'}$ is the reward expectation of the optimal algorithm.
\end{theorem}
\begin{IEEEproof}
Algorithm~\ref{alg:mab scheme} can operate at the time slots in the set $\period_a$, where $|\period_a|\leq|\period'|$.
Suppose that $\period_a=\period_1\cup\period_2$. In the period $\period_1$, the arm $\arm_j$, $j=1,2,3$, are pulled, and in the period $\period_2$, the arm $\arm_4$ is pulled.
Suppose that the optimal algorithm operates at the time slots in the set $\period^*$. $|\period^*|\leq|\period'|$ and $|\period_a|\leq|\period'|$ because the charging efficiency $\charge<1$.
The reward regret of \protocol is:
\begin{align}
&\regret_{\protocol}(\Energy)=\expect[\reward_{\scheme^*}(\Energy)]-\expect[\reward_{\protocol}(\Energy)] \nonumber\\
&=\expect[\sum\limits^\period_{\slot=1}\reward^*(\slot)]-\expect[\sum\limits^\period_{\slot=1}\reward(\slot)] = \expect[\sum\limits^{\period^*}_{\slot=1}\reward^*(\slot)]-\expect[\sum\limits^{\period_a}_{\slot=1}\reward(\slot)]\nonumber\\
&=\expect[\sum\limits^{\period_1}_{\slot=1}\sum\limits^{\armnum-1}_{j=1}(\reward^*(\slot)-\reward_j(\slot))]
 +\expect[\sum\limits^{\period_2}_{\slot=1}(\reward^*(\slot)-\reward_4(\slot))]\nonumber\\
 &\hspace{2cm}+\expect[\sum\limits^{\period^*-\period_1-\period_2}_{\slot=1}\reward^*(\slot)]\nonumber\\
 &\leq\expect[\sum\limits^{\period_1}_{\slot=1}\sum\limits^{\armnum-1}_{j=1}(\reward^*(\slot)-\reward_j(\slot))]
 +\expect[\sum\limits^{\period^*-\period_1}_{\slot=1}(\reward^*(\slot)-\reward_4(\slot))]\nonumber\\
 &=\nonumber\expect[\sum\limits^{\period_1}_{\slot=1}\sum\limits^{\armnum-1}_{j=1}\rewardd_j\prob(\arm(\slot)=\arm_j)]
 +\expect[\sum\limits^{\period^*-\period_1}_{\slot=1}\reward^*(\slot)\pullarmnumber_4(\slot)]\\
 &=\nonumber\expect[\sum\limits^{\period_1}_{\slot=1}\sum\limits^{\armnum-1}_{j=1}\rewardd_j\pullarmnumber_j(\slot)]
 +\expect[\sum\limits^{\period^*-\period_1}_{\slot=1}\reward^*(\slot)\pullarmnumber_4(\slot)]\\
&<\sum^{\armnum-1}\limits_{j=1}[(\frac{\cost_{max}}{\cost_{min}})^2\frac{\parae'\ln|\period'|}{\rewardd_j}+2\rewardd_j]\nonumber\\
&\hspace{2cm}+u^{*'}[(\frac{\cost_{max}}{\cost_{min}})^2\frac{\parae'\ln|\period'|}{\rewardd_4^2}+1]\nonumber
\end{align}
where $\arm(\slot)$ denotes the arm pulled at $\slot$, and the reward of the arm $\arm_4$ is $\reward_4$, which is zero since to store energy cannot process data.
This finishes the proof.
\end{IEEEproof}
\section{Experiment Results}
\label{sec:experiment}
This section depicts our experiments established on the real data obtained from the real solar harvesting module: SolarMote~\cite{shen2009solarmote}.
A series of experiments are designed and  implemented to validate the performance of our scheme \protocol by comparing with two baseline algorithms: \schemeO and \schemeS, which are designed
because of the hardness to find the optimal algorithm for the opportunistic duty cycling.
The strong assumption behind \schemeO is that the VoI of the data to process and the harvested energy can be previously known while no extra energy is consumed on the energy storage. \schemeO is a centralized and off-line algorithm.
Thus, the performance of \schemeO should be  closer to the optimal algorithm than \protocol and \schemeS.
\schemeS predicts the amount of the energy to harvest and then calculates the duty cycle in advance as the typical algorithm given in the reference~\cite{jiang2005perpetual}.
In the following context,  two scenarios: single sensor node and network, are established to evaluate the performance of these algorithms.
For the algorithm \protocol and \schemeS,  the charge efficiency $\charge=80\%$.
The time slot $\slot$ is set to be 60 seconds. The energy threshold is set to be 20mA.
All experiments in this section are simulated on the network simulation platform OMNeT++ 4.1~\cite{omnetpp}.
\subsection{Single Node Scenario}
This subsection simulates the scenario consisting of only one sensor node $\node_1$ and the sink.
$\node_1$  samples data from its surrounding, and transmits its data to the sink.
The scenario contains four experiments and is set up to evaluate the impact of the chance to harvest energy by excluding the impact of other factors occurring in the large scale networks, such as the packet loss.
Each experiment evaluates the reward performance of the three algorithms: \schemeO, \protocol and \schemeS.

\begin{figure}\centering
\begin{minipage}[b]{.22\textwidth}
  \includegraphics[scale=0.57,bb=115 624 319 781]{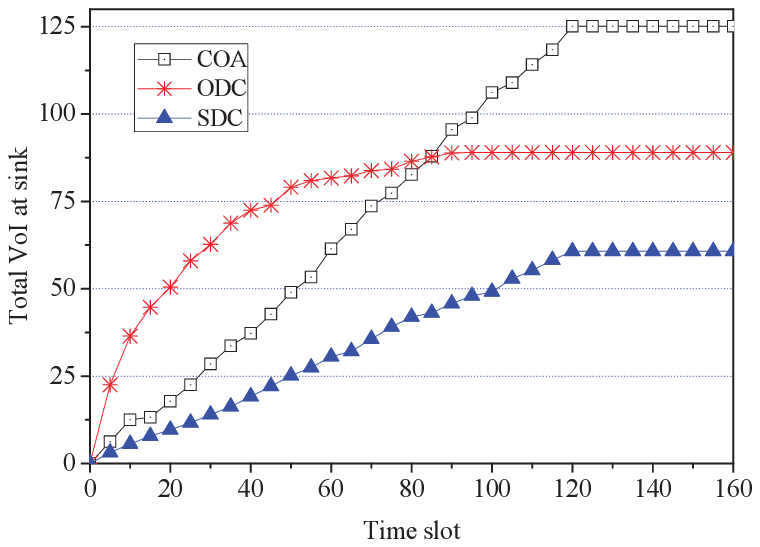}
  \caption{\label{fig:information value derived with fixed energy}The fixed energy is assigned to one phase.}
\end{minipage}
\hspace{0.2cm}\begin{minipage}[b]{.22\textwidth}
  \includegraphics[scale=0.57,bb=113 625 318 774]{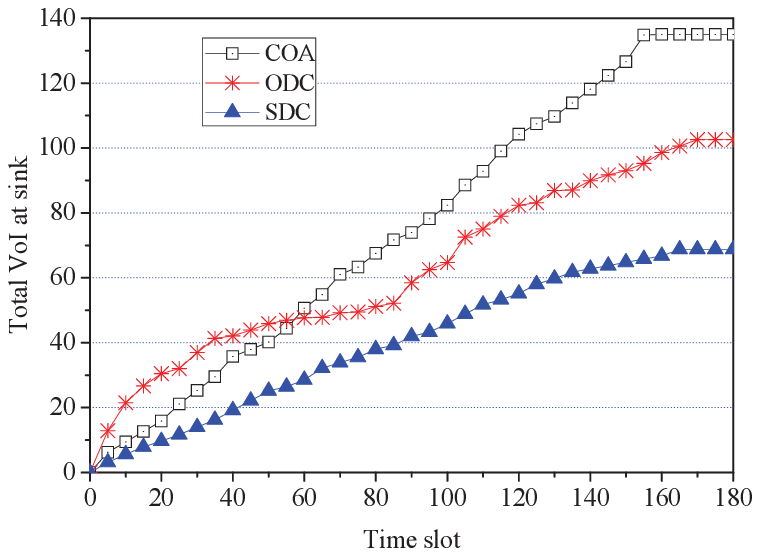}
  \caption{\label{fig:information value with two fixed energy}The fixed energy is assigned equally to two phases.}
\end{minipage}
\end{figure}

\begin{figure}\centering
\begin{minipage}[b]{.22\textwidth}
  \includegraphics[scale=0.54,bb=110 622 319 776]{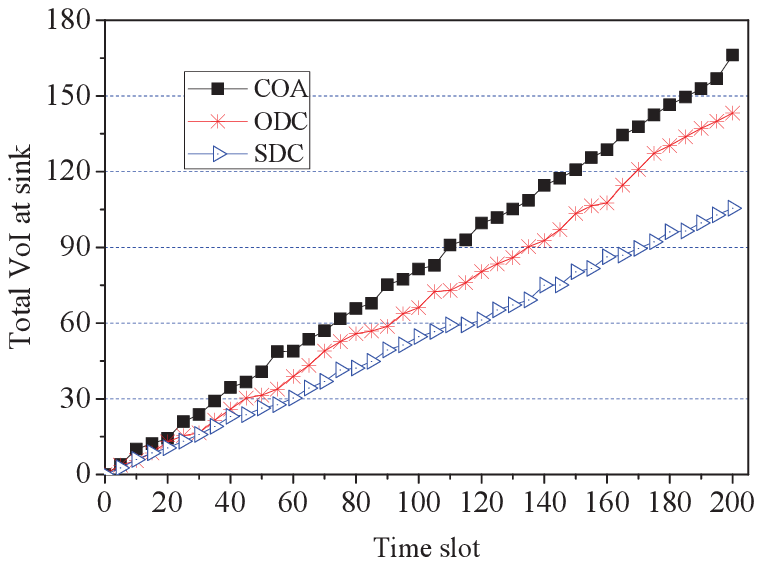}
  \caption{\label{fig:assumed information value derived by one node}Uniformly and randomly distributed data and energy.}
\end{minipage}
\hspace{0.3cm}\begin{minipage}[b]{.22\textwidth}
  \includegraphics[scale=0.54,bb=109 622 317 775]{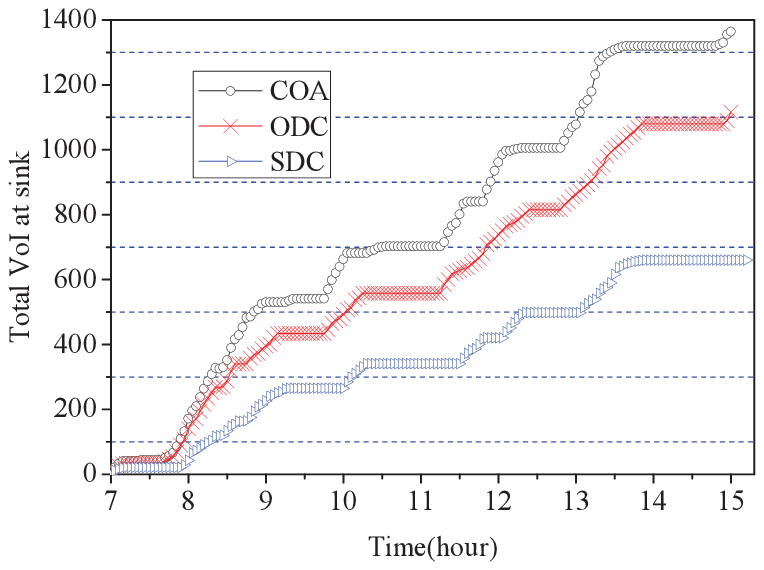}
  \caption{\label{fig:information value derived by node 2 with real data}The luminous intensity and harvested energy measured by SolarMote.}
\end{minipage}
\end{figure}

In the first experiment, the fixed amount 1mAh of energy is  previously assigned in the  phase from the time slots 0 to 10.
In the second experiment, the 1mAh energy is divided into two equivalent parts.
One part is assigned to the sensor node in the phase from the time slots 0 to 5 while the other is assigned in the phase from the time slots 90 to 95.
In the third experiment, the energy 1mAh is divided into 180 units, which are uniformly and randomly distributed into the period from the slot 0 to 200.
In the three experiments, there is one data available in each time slot, and its VoI is assumed to follow the Gaussian probability distribution with the expectation 1 and the variance 0.5.
In the fourth experiment, the data to process  and the energy  to harvest are  the real data collected by the energy harvesting module SolarMote: the luminous intensity and harvested energy in the first sub-figure of Figure~\ref{subfig:Several nodes in one day}.
Assume that there is 20mAh initial energy in the sensor node's battery.
The simulations for each of the experiments are  repeatedly run for 100 times so each data point in Figure~\ref{fig:information value derived with fixed energy}$\scriptsize{\sim}$\ref{fig:information value derived by node 2 with real data} is the average of this amount of trials.

The results of the first and second experiments are respectively illustrated in Figure~\ref{fig:information value derived with fixed energy} and \ref{fig:information value with two fixed energy}.
These experiment results  indicate the impact of the energy harvesting access on the total VoI.
When the energy is assigned at the fixed phases, the sensor nodes tend to spend the energy timely at these phases  by \protocol since some extra energy must be consumed to store the energy.
In Figure~\ref{fig:information value derived with fixed energy}, the growth rate of $\node_1$'s total VoI by \protocol is higher than those of \schemeO and \schemeS at the initial phase.
Although the total VoI under \protocol slows down its growth in the first experiment in Figure~\ref{fig:information value derived with fixed energy} after the initial phase,
the finally total VoI of \protocol is 28.25\% higher than \schemeS, and 28.86\% lower than that of \schemeO.
In the second experiment, the fixed energy is assigned at two phases.
At the two phases, the growth rate of total VoI by \protocol suddenly increases since the two phases are considered to be good chance to use the energy by \protocol.
The finally total VoI of \protocol 34.62\% higher than \schemeS, and 24.02\% lower that of \schemeO as shown in Figure~\ref{fig:information value with two fixed energy}.

The results of the third experiment is illustrated in Figure~\ref{fig:assumed information value derived by one node}.
The total VoI of all algorithms grows almost linearly with time.
Till the time slot 200,  the total VoI obtained by \schemeO, \protocol and \schemeS  are 166.14, 143.195 and 105.509 respectively.
\protocol is 16.02\% lower than   \schemeO, and  35.72\% higher than \schemeS.
Because the fixed energy 1mAh is distributed in the whole period from the slot 0 to 200 uniformly, there are much more chances that each sensor node need not store the harvested energy but to consume it directly. The three algorithms thus can obtain much more VoI than those in the first and second experiments.
 
In the fourth experiment,  we adopt the real data harvested by SolarMote~\cite{shen2009solarmote} including the  luminous intensity and the harvested energy  as shown in the first subgraph of Figure~\ref{subfig:Several nodes in one day}.
Thus, the data that the sensor node $\node_1$ will process is the luminous intensity.
In each time slot, $\node_1$ can harvest  different amount of energy and luminous intensity, and cannot know the exact information of the luminous intensity in the future time slots.
As shown in Figure~\ref{fig:information value derived by node 2 with real data}, the experiment results illustrate that the finally total VoI of \protocol is 18.18\% lower than that of \schemeO, and 69.09\% higher than that of \schemeS.
Different from the previous experiments, the accesses to harvest energy and  the data are inhomogeneous over time.
Compared to the results in the third experiment, the performance of \protocol is better than \schemeS.
Because of the inhomogeneous accesses, \protocol loses some better chances to process the data and its performance decreases compared to the result in the third experiment.

In the four experiments, the occasions that the sensor node $\node_1$ can harvest energy increase from one short phase as the first experiment to multiple moments as the third and fourth ones.
In the third experiment, $\node_1$ has uniformly possibility to harvest energy at each time slot while having the heterogeneous possibility in the fourth experiment.
Through the four experiments, we find that the distribution of the accesses to harvest energy and to process data have much impact on the performance of  the duty cycling scheme.
The overall performance of \protocol is  closer to that of the centralized algorithm \schemeO than \schemeS when the access to harvest energy  and to process data distribute more uniformly.
In the four experiments, the reward regret of \protocol are given by comparing to the performance of \schemeO because of the hardness to find the optimal scheme. The numerical values of the regret are presented in the form of the percentage in the above analysis.
\subsection{Network Scenario}
\begin{figure}\centering
\begin{minipage}[b]{.22\textwidth}
  \includegraphics[scale=0.57,bb=117 617 318 774]{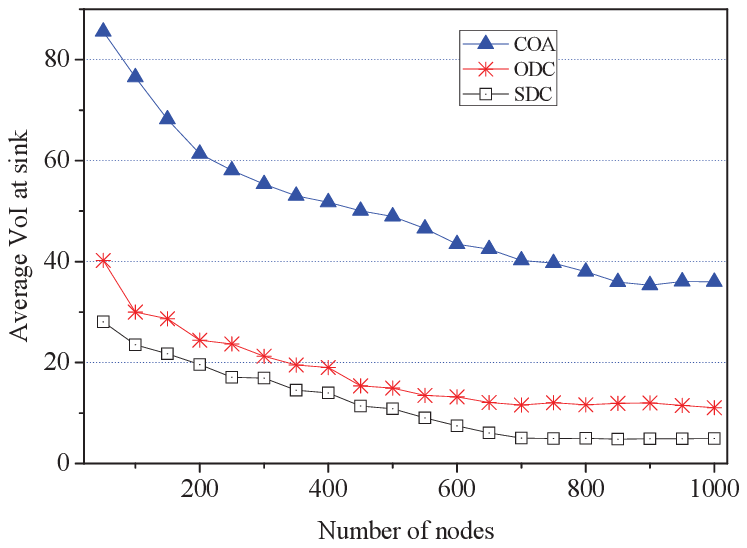}
  \caption{\label{fig:node density fixed energy}Fixed energy is assigned at the initial phases of each sensor node.}
\end{minipage}
\hspace{0.4cm}\begin{minipage}[b]{.22\textwidth}
  \includegraphics[scale=0.57,bb=110 617 319 774]{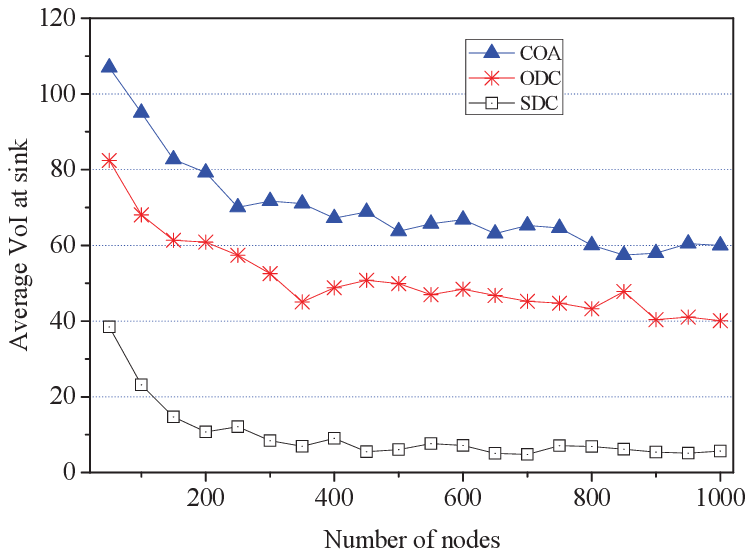}
  \caption{\label{fig:node desity allocated energy} Energy is randomly allocated to each time slot.}
\end{minipage}
\end{figure}

The node density has also obvious impact on the performance of these algorithms.
This block simulates the network scenario  composed of a several numbers of the sensor nodes with the fixed deploy area of 100$\times$100 $m^2$.
Each sensor node runs the IEEE 802.15.4 protocol to assign wireless channel and deal with the interference among the sensor nodes.
All sensor nodes have the same receiving and minimal transmitting range as far as 50 meters.
Each data point  represents the average  VoI when the network scale is a specific number of the sensor nodes.
 For example, the most left data point in Figure~\ref{fig:node density fixed energy} represents the average VoI when the network has 50 sensor nodes.
The simulation time  is set to be 200 time slots and every slot is one minute.
Two experiments are designed and implemented.
In the first one, the energy 1mAh is assigned to each sensor node at the initial phase from the time slot 0 to 10.
In the second one,  the energy 1mAh is divided equally into 180 units and distributed to the 200 time slots uniformly and randomly.
The simulation results of the two experiments are respectively shown in Figure~\ref{fig:node density fixed energy}
    and \ref{fig:node desity allocated energy}.
In these figures, each data point is the average of 10 trials.

From the results of both experiments in Figure~\ref{fig:node density fixed energy} and~\ref{fig:node desity allocated energy}, it is easy to notice that the increase of the node density results in the much decrease of VoI per sensor node.
In the first experiment, the average VoI under the algorithms: \schemeO, \protocol and \schemeS drop 82.35\%, 72.54\% and 57.92\% respectively when the number of the sensor  nodes increases from 50 to 1000.
Similarly, the average VoI drops 82.35\%, 48.95\% and 43.897\% respectively in the second experiment.
In spite of the VoI degression with the increasing of the node density,  the performance of \protocol is over that of the \schemeS.
The average VoI of \protocol can be 47. 01\% of \schemeO while \schemeS is at most 32.08\% of \schemeO in the first experiment.
 The average VoI  of \protocol is at least 67.89\% of that by \schemeO while \schemeS is at most 35.96\% of \schemeO in the second experiment.

The section evaluates the performance of our algorithm \protocol  by comparing with the centralized algorithm \schemeO and the algorithm \schemeS representing the typical duty cycling scheme.
 The results show that the \protocol  has stable performance over the existing scheme.
 Although its performance is not as good as the centralized algorithm \schemeO, \protocol has acceptable performance as a distributed   online algorithm.
 Through the experiments in the section, we found that the access to process data and to harvest energy have much impact on the performance of the duty cycling scheme, and thus the energy efficiency.
  \protocol, as the first work, shows the impact under the requirement of the energy neutral operation, and has promising performance.
\section{Related Work}
\label{sec:related}
This section reviews the existing energy harvesting modules and the duty cycling schemes in the energy harvesting WSNs.
Energy harvesting technique and applications for WSNs have been widely studied~\cite{he2012energy}\cite{he2013energy}\cite{zhang2013data} and they designed different modules to harvest energy.

\textbf{Energy harvesting module.}
Some typical modules, such as Prometheus~\cite{jiang2005perpetual} and Helimote~\cite{hsu2005energy}, were designed for the sensor node  to harvest solar,
         vibration and wind energy~\cite{liu2011perpetual}\cite{ozel2012achieving}.
Most of existing modules can harvest the solar energy by the micro-scale photovoltaic power system.
Actually, duty cycling is still necessary in the energy harvesting WSNs because some works~\cite{jiang2005perpetual}\cite{gu2009esc} found that nodes had not intensive enough energy to sustain the continuous full duty cycle operation of sensor nodes yet although they could harvest ambient energy.
For example, the solar panel in the module Prometheus requires at least 4 hour hard light each day if the node's duty cycle is 10\%~\cite{jiang2005perpetual}.
Some modules were designed to store the harvested energy into their batteries while others took capacitor as the primary buffer and battery as the second buffer~\cite{jiang2005perpetual}.
Battery suffers from low charge efficiency and long charging duration,  capacitor has the high leakage~\cite{zhu2009leakage}.
For example, the 2000F ultra-capacitor has the high leakage rate up to 43.8\% during the first month~\cite{zhu2009leakage}.

Although the sensor network can obtain the environmental energy from time to time, the harvested energy is not enough to support full duty cycle.
This paper notices the phenomenon and argues to spend the harvested energy on the proper time so that the energy efficiency can be improved while each node has enough energy to support the energy neutral operation.

\textbf{Duty cycling.}
Duty cycling has been constantly researched as a promising technique to improve energy efficiency and prolong network life because of
      the energy limitation~\cite{ghidini2011energy}\cite{li2012fairness}\cite{xu2015low}.
We can group these previous techniques into two classes: classical and adaptive duty cycling, according to the way under which the networks are powered.
In WSNs, the classical duty cycling based on the preliminary assumption that each sensor has limited energy, \ie, no extra energy is supplemented~\cite{li2012fairness}\cite{guo2011correlated}\cite{tang2011cool}.
So the goal of the classical duty cycling is to save energy as much as possible.
In the energy harvesting WSNs, each sensor node can be supplemented extra energy continuously by the energy harvesting modules~\cite{kansal2007power}\cite{vigorito2007adaptive}.
Only a few works were engaged in adjusting duty cycle according to the weather conditions in order to achieve
      high energy efficiency~\cite{kansal2007power}\cite{vigorito2007adaptive}\cite{Buchli2014sensys}.
However, they did not consider that the energy harvesting is a random process~\cite{ho2010markovian}, and the energy profile of each sensor node is time-varying and different from others~\cite{kansal2004performance}\cite{zhu2009leakage}.
Furthermore, these previous works adjusted duty cycle only according to the predicted amount of the harvested energy in a duration, and did not consider the impact of network demand and the occasion to implement the demand on duty cycling.

Different from the previous works, this paper addresses the duty cycling with the dynamic harvested energy to avoid the complex prediction or rough estimation of the duty cycle.
We consider the spatiotemporal dynamic of the harvested energy and our scheme ensures that each sensor node adjusts its duty cycle according to its local information on the energy harvesting and data process.
\section{Conclusion}
\label{sec:conclusion}
This paper investigates the process of  the energy harvesting in the solar sensor networks, and finds the phenomenon that it can greatly improve the energy efficiency of the harvested energy to catch right chance to use it.
We formulate  it as the budget-dynamic MAB problem, and propose  the new scheme: \protocol, to exploit the harvested energy fully while satisfying the energy neutral operation.
The theoretical   performance for the scheme are analyzed, and the experimental analysis is also designed and implemented.
We are the first to study the phenomenon, and will go on studying the problem in the following aspects.
Because of the simplicity of the sensor node hardware, it needs more simple and feasible schemes.
Furthermore, the scheme of this paper involves some frequent update the probability to choose arms.
Our future work will reduce the computation frequency and design even simple scheme requiring a few frequency to update the  probability.
Thirdly, the synchronization and communication coordination among nodes are quite helpful to improve the channel utilization.
In the coming work, we will consider the coordination among sensor nodes.


\bibliographystyle{unsrt}
\bibliography{pakeyref}

\begin{thebibliography}{10}

\bibitem{Mainwaring2002}
Alan Mainwaring, David Culler, Joseph Polastre, Robert Szewczyk, and John
  Anderson.
\newblock Wireless sensor networks for habitat monitoring.
\newblock In {\em Proceedings of the 1st ACM International Workshop on Wireless
  Sensor Networks and Applications (WASA)}, pages 88--97. ACM, 2002.

\bibitem{shen2009solarmote}
Xingfa Shen, Cheng Bo, Jianhui Zhang, and et~al.
\newblock {SolarMote}: a low-cost solar energy supplying and monitoring system
  for wireless sensor networks. poster.
\newblock In {\em Proceeding of the 7th Conference on Embedded Networked Sensor
  Systems (SenSys)}, pages 413--414, Berkeley, California, USA, Nov. 4-6 2009.
  ACM.

\bibitem{tang2011cool}
S.J. Tang, X.Y. Li, X.~Shen, Jianhui Zhang, G.~Dai, and S.K. Das.
\newblock Cool: On coverage with solar-powered sensors.
\newblock In {\em Proceedings of the 31st International Conference on
  Distributed Computing Systems (ICDCS)}, pages 488--496, Minneapolis,
  Minnesota, USA, June 20-24 2011. IEEE.

\bibitem{jiang2005perpetual}
X.~Jiang, J.~Polastre, and D.~Culler.
\newblock Perpetual environmentally powered sensor networks.
\newblock In {\em Proceeding of the 4th International Conference on Information
  Processing in Sensor Networks (IPSN)}, pages 463--468, (UCLA) Los Angeles,
  CA, USA, April 25-27 2005. ACM/IEEE.

\bibitem{ghidini2011energy}
G.~Ghidini and S.K. Das.
\newblock An energy-efficient markov chain-based randomized duty cycling scheme
  for wireless sensor networks.
\newblock In {\em Proceedings of the 31st International Conference on
  Distributed Computing Systems (ICDCS)}, pages 67--76, Minneapolis, Minnesota,
  USA, June 20-24 2011. IEEE.

\bibitem{gu2009esc}
Y.~Gu, T.~Zhu, and T.~He.
\newblock {ESC}: Energy synchronized communication in sustainable sensor
  networks.
\newblock In {\em Proceeding of the 17th International Conference on Network
  Protocols (ICNP)}, pages 52--62, Princeton, New Jersey, USA, Oct. 13-16 2009.
  IEEE.

\bibitem{Shen2013EFCon}
Jianhui Zhang Shaojie Tang Xufei Mao Guojun~Dai Xingfa~Shen, Cheng~Bo.
\newblock {EFC}on: Energy flow control for sustainable wireless sensor
  networks.
\newblock {\em Elsevier Ad Hoc Networks}, 11(4):1421¨C1431, 2013.

\bibitem{kansal2007power}
A.~Kansal, J.~Hsu, S.~Zahedi, and M.B. Srivastava.
\newblock Power management in energy harvesting sensor networks.
\newblock {\em ACM Transactions on Embedded Computing Systems}, 6(4):32, 2007.

\bibitem{moser2010adaptive}
C.~Moser, L.~Thiele, D.~Brunelli, and L.~Benini.
\newblock Adaptive power management for environmentally powered systems.
\newblock {\em IEEE Transactions on Computers}, 59(4):478--491, 2010.

\bibitem{Buchli2014sensys}
Bernhard Buchli, Felix Sutton, Jan Beutel, and Lothar Thiele.
\newblock Dynamic power management for long-term energy neutral operation of
  solar energy harvesting systems.
\newblock In {\em Proceeding of the 12th Conference on Embedded Networked
  Sensor Systems (SenSys)}, pages 31--45, Memphis, TN, USA, Nov. 3-6 2014. ACM.

\bibitem{ding2000battery}
Y.~Ding, R.~Michelson, and C.~Stancil.
\newblock Battery state of charge detector with rapid charging capability and
  method, July~25 2000.
\newblock US Patent 6,094,033.

\bibitem{zhu2009leakage}
T.~Zhu, Z.~Zhong, Y.~Gu, T.~He, and Z.L. Zhang.
\newblock Leakage-aware energy synchronization for wireless sensor networks.
\newblock In {\em Proceeding of the 7th Annual International Conference on
  Mobile Systems, Applications, and Services (MobiSys)}, pages 319--332,
  Krak¨®w, Poland, June 22-25 2009. ACM.

\bibitem{ho2010markovian}
C.K. Ho, P.D. Khoa, and P.C. Ming.
\newblock Markovian models for harvested energy in wireless communications.
\newblock In {\em Proceeding of the International Conference on Communication
  Systems (ICCS)}, pages 311--315. ACM, 2010.

\bibitem{Liu2010Distributed}
Qing~Zhao Keqin~Liu.
\newblock Distributed learning in multi-armed bandit with multiple players.
\newblock {\em IEEE Transactions on Signal Processing}, 58(11):5667 -- 5681,
  2010.

\bibitem{kansal2004performance}
A.~Kansal, D.~Potter, and M.B. Srivastava.
\newblock Performance aware tasking for environmentally powered sensor
  networks.
\newblock In {\em ACM SIGMETRICS Performance Evaluation Review}, volume~32,
  pages 223--234, 2004.

\bibitem{park2006ambimax}
C.~Park and P.H. Chou.
\newblock Ambimax: Autonomous energy harvesting platform for multi-supply
  wireless sensor nodes.
\newblock In {\em Proceeding of the 3rd Annual Communications Society on Sensor
  and Ad Hoc Communications and Networks (SECON)}, volume~1, pages 168--177,
  Hyatt Regency, Reston, VA, USA, Sep. 25 - 28 2006. IEEE.

\bibitem{padhy2010utility}
P.~Padhy, R.K. Dash, K.~Martinez, and N.R. Jennings.
\newblock A utility-based adaptive sensing and multihop communication protocol
  for wireless sensor networks.
\newblock {\em ACM Transactions on Sensor Networks}, 6(3):27, 2010.

\bibitem{ventura2011markov}
J.~Ventura and K.~Chowdhury.
\newblock Markov modeling of energy harvesting body sensor networks.
\newblock In {\em Proceeding of the 22nd International Symposium on Personal
  Indoor and Mobile Radio Communications (PIMRC)}, pages 2168--2172, Toronto,
  Canada, Sept. 11-14 2011. IEEE.

\bibitem{gittins2011multi}
J.~Gittins, K.~Glazebrook, and R.~Weber.
\newblock {\em Multi-armed bandit allocation indices}.
\newblock Wiley, 2011.

\bibitem{smallwood1973optimal}
Richard~D Smallwood and Edward~J Sondik.
\newblock The optimal control of partially observable markov processes over a
  finite horizon.
\newblock {\em Operations Research}, 21(5):1071--1088, 1973.

\bibitem{kohli2004average}
Rajeev Kohli, Ramesh Krishnamurti, and Prakash Mirchandani.
\newblock Average performance of greedy heuristics for the integer knapsack
  problem.
\newblock {\em European Journal of Operational Research}, 154(1):36--45, 2004.

\bibitem{auer2002finite}
Peter Auer, Nicol{\`o} Cesa-Bianchi, and Paul Fischer.
\newblock Finite-time analysis of the multiarmed bandit problem.
\newblock {\em Machine learning}, 47(2):235--256, 2002.

\bibitem{kumar1986stochastic}
Panqanamala~Ramana Kumar and Pravin Varaiya.
\newblock {\em Stochastic systems: estimation, identification and adaptive
  control}.
\newblock Prentice-Hall, Inc., 1986.

\bibitem{omnetpp}
{OMN}e{T}++.
\newblock http://www.omnetpp.org/.
\newblock Accessed: Nov.22$^{nd}$, 2014.

\bibitem{he2012energy}
Shibo He, Jiming Chen, David~KY Yau, Huanyu Shao, and Youxian Sun.
\newblock Energy-efficient capture of stochastic events under periodic network
  coverage and coordinated sleep.
\newblock {\em IEEE Transactions on Parallel and Distributed Systems},
  23(6):1090--1102, 2012.

\bibitem{he2013energy}
Shibo He, Jiming Chen, Fachang Jiang, David~KY Yau, Guoliang Xing, and Youxian
  Sun.
\newblock Energy provisioning in wireless rechargeable sensor networks.
\newblock {\em IEEE Transactions on Mobile Computing}, 12(10):1931--1942, 2013.

\bibitem{zhang2013data}
Yongmin Zhang, Shibo He, and Jiming Chen.
\newblock Data gathering optimization by dynamic sensing and routing in
  rechargeable sensor networks.
\newblock In {\em Proceeding of the 10th Annual IEEE Communications Society
  Conference on Sensor, Mesh and Ad Hoc Communications and Networks (SECON)},
  pages 273--281, New Orleans, USA, June 24-27 2013. IEEE.

\bibitem{hsu2005energy}
J.~Hsu, A.~Kansal, J.~Friedman, V.~Raghunathan, and M.~Srivastava.
\newblock Energy harvesting support for sensor networks.
\newblock {\em the Sensor Platforms, Tools and Design Methods (SPOTS) track at
  the 4th International Conference on Information Processing in Sensor Networks
  (IPSN)}, April 25-27 2005.

\bibitem{liu2011perpetual}
R.S. Liu, K.W. Fan, Z.~Zheng, and P.~Sinha.
\newblock Perpetual and fair data collection for environmental energy
  harvesting sensor networks.
\newblock {\em IEEE/ACM Transactions on Networking}, 19(4):947--960, 2011.

\bibitem{ozel2012achieving}
Omur Ozel and Sennur Ulukus.
\newblock Achieving {AWGN} capacity under stochastic energy harvesting.
\newblock {\em IEEE Transactions on Information Theory}, 58(10):6471--6483,
  2012.

\bibitem{li2012fairness}
Zhenjiang Li, Mo~li, and Youhao Liu.
\newblock Towards energy-fairness in asynchronous duty-cycling sensor networks.
\newblock In {\em Proceeding of the 31st Annual IEEE International Conference
  on Computer Communications (INFOCOM)}, pages 801--809, Orlando, Florida USA,
  March 25-30 2012. IEEE.

\bibitem{xu2015low}
Lijie Xu, Haipeng Dai, and Guihai Chen.
\newblock A low--delay energy--efficient decision strategy in duty--cycled
  wireless sensor networks.
\newblock {\em International Journal of Sensor Networks}, 17(2):82--92, 2015.

\bibitem{guo2011correlated}
S.~Guo, S.M. Kim, T.~Zhu, Y.~Gu, and T.~He.
\newblock Correlated flooding in low-duty-cycle wireless sensor networks.
\newblock In {\em Proceeding of the 19th International Conference on Network
  Protocols (ICNP)}, pages 383--392, Vancouver, Canada, October 17-20 2011.
  IEEE.

\bibitem{vigorito2007adaptive}
C.M. Vigorito, D.~Ganesan, and A.G. Barto.
\newblock Adaptive control of duty cycling in energy-harvesting wireless sensor
  networks.
\newblock In {\em Proceeding of the 4th Annual Communications Society
  Conference on Sensor, Mesh and Ad Hoc Communications and Networks (SECON)},
  pages 21--30, San Diego, California, USA, June 18-21 2007. IEEE.

\end{thebibliography}
\end{document}